\documentclass[journal,twoside,web]{ieeecolor}
\usepackage{generic}
\usepackage{placeins} 
\usepackage[utf8]{inputenc}
\usepackage{cite}
\usepackage{stfloats}
\usepackage{amsmath,amssymb,amsfonts}
\usepackage{balance}
\allowdisplaybreaks
\usepackage{algorithmic}
\usepackage{booktabs}
\usepackage{multirow}
\usepackage{mathtools}
\usepackage{array}
\usepackage{graphicx}
\usepackage[caption=false,font=footnotesize]{subfig} 
\usepackage{gensymb}
\usepackage{algorithm,algorithmic}
\newtheorem{assumption}{Assumption}
\newtheorem{lemma}{Lemma}
\newtheorem{remark}{Remark}

\newtheorem{proposition}{Proposition}
\newtheorem{theorem}{Theorem}
\usepackage{comment}
\usepackage{xurl}
\usepackage{hyperref}
\hypersetup{hidelinks=true}
\usepackage{textcomp}
\usepackage{tikz}
\usetikzlibrary{3d,perspective}

\newenvironment{hw}
  {\par\color{red}\textbf{HW:}~}
  {\par}

\tikzset
{%
  pics/cylinder/.style n args={3}{
    code={%
      \draw[pic actions] (135:#1) arc (135:315:#1) --++ (0,0,#2) arc (315:135:#1) -- cycle;
      \draw[pic actions] (0,0,#2) circle (#1);
      \foreach\z in {0,1}
      {
        \begin{scope}[canvas is xy plane at z=\z*\h]
          \coordinate (-cen\z) at       (0,0);
          \coordinate (-ESE\z) at    (-#3:#1);
          \coordinate (-ENE\z) at     (#3:#1);
          \coordinate (-NNE\z) at  (90-#3:#1);
          \coordinate (-NNW\z) at  (90+#3:#1);
          \coordinate (-WNW\z) at (180-#3:#1);
          \coordinate (-WSW\z) at (180+#3:#1);
          \coordinate (-SSW\z) at (270-#3:#1);
          \coordinate (-SSE\z) at (270+#3:#1);
        \end{scope}
      }
    }},
}
\def\BibTeX{{\rm B\kern-.05em{\sc i\kern-.025em b}\kern-.08em
    T\kern-.1667em\lower.7ex\hbox{E}\kern-.125emX}}
\begin{document}
\title{Model Predictive Control with Multiple Constraint Horizons}
\author{Allan Andre do Nascimento, Han Wang, Antonis Papachristodoulou, \IEEEmembership{Fellow, IEEE}, Kostas Margellos \IEEEmembership{Senior Member, IEEE}
\thanks{Submission for review: dd/mm/yyyy. AAdN, AP and KM acknowledge funding support by MathWorks. AP was supported in part by UK’s Engineering and Physical Sciences Research Council projects EP/X017982/1, EP/Y014073/1 and UKRI2108. For the purpose of Open Access, the authors have applied a CC BY public copyright licence to any Author Accepted Manuscript (AAM) version arising from this submission. }
\thanks{Allan Andre do Nascimento, Antonis Papachristodoulou and Kostas Margellos are with the Department of Engineering Science, University of Oxford,  Parks Road, Oxford OX1 3PJ, UK (email: \{allan.adn, antonis, kostas.margellos \}@eng.ox.ac.uk).}
\thanks{Han Wang is with the Department of Information Technology and Electrical Engineering, ETH Zurich,  ETL, Physikstrasse 3, 8092 Z{\"u}rich, Switzerland (email: hanwang@control.ee.ethz.ch).}}

\maketitle

\begin{abstract}

\textcolor{black}{We propose a Model Predictive Control (MPC) formulation for nonlinear systems without terminal penalty or dedicated stabilizing terminal set, in which state constraints are enforced heterogeneously  along the prediction horizon. In this setting a control-invariant set certifies near-term safety, while a less restrictive set constrains later predictions. This structure is motivated by safety-critical formulations, such as Control Barrier Function (CBF) based MPC, collision avoidance, and robotic receding-horizon planning, where near-term predictions must be certified safe and later predictions can be re-certified in future updates. We develop a value-function-difference analysis separating the effects of constraint-set selection and prediction-horizon length on closed-loop performance bounds, yielding implicit suboptimality certificates. Assuming cost-controllability, we further propose an upper-bound certificate which accounts explicitly for distinct constraint sets, their associated decay rates, and horizons. We also provide a lower-bound certificate for the closed-loop cost beyond the finite-horizon open-loop cost, which, for the admissible parameter range, is not weaker than the standard finite-horizon bound. Simulations on linear and nonlinear safety-critical systems demonstrate the proposed  certificates \textit{a priori} and \textit{a posteriori}.
}
\end{abstract}


\section{Introduction}
\label{sec:introduction}
\IEEEPARstart{M}{odel} Predictive Control (MPC) \cite{grne2013nonlinear} is an optimization-based method that approximates the infinite-horizon constrained problem by a finite-horizon one \cite{grune2007computing}, raising the question of how similar these two \textcolor{black}{solutions} are. This can be addressed using closed-loop suboptimality analysis \cite{grune2017nonlinear}.

An early contribution on closed-loop optimality analysis is \cite{shamma1997linear}, where bounds on closed-loop suboptimality for linear discrete-time systems were derived, by assuming the availability of finite-horizon optimal value functions. \textcolor{black}{For nonlinear MPC without terminal ingredients, stability was studied in \cite{grimm2005model} without requiring a local control Lyapunov terminal cost, in \cite{tuna2006shorter} with horizon-related refinements and in \cite{grimm2007nominally} with robustness and state-constraint aspects.} Nonlinear system investigations were \textcolor{black}{pursued} in \cite{grune2008infinite}, where open-loop value functions were used to evaluate closed-loop suboptimality, yielding a sufficient stability condition \textcolor{black}{as a function of} the prediction horizon. Previous literature also focused on the prediction horizon as the tuning dial for suboptimality estimation. Suboptimality was analyzed in \cite{grune2009practical} for open-loop costs upper-bounded by a chosen running cost within the horizon, assuming uniform constraints and access to optimal state--input pairs. \textcolor{black}{Subsequently } \cite{worthmann2011stability,grune2017nonlinear,grune2018dynamic} moved from uniform to dynamic ratios between the open-loop value function and running cost, enabling tighter suboptimality bounds by capturing changes in open and closed-loop costs with greater precision. Extending the knowledge of MPC parameters' effect on suboptimality, \cite{worthmann2011stability} studied the effect of using a control horizon larger than one in closed-loop, relevant to networked MPC, with communication delays and data losses \cite{findeisen2011robustness}. Suboptimality was also studied for economic MPC (e.g., \cite{grune2013economic,faulwasser2018economic}).

Recent developments in safety-critical control have sparked a growing interest in enforcing set invariance \cite{blanchini1999set}, particularly through the Control Barrier Function (CBF) framework \cite{wieland2007constructive, ames2019control}. This form of constraint enforcement is especially well-suited to receding horizon techniques, which are themselves advantageous in safety-critical problems due to their predictive capabilities \cite{wabersich2021predictive}. Consequently, MPC-like controllers that integrate CBF constraints have gained traction, with efforts spanning centralized \cite{ zeng2021safety}, distributed \cite{do2023game}, probabilistic \cite{do2024probabilistically}, learning \cite{sabouni2024reinforcement}, and real-time \cite{grandia2021multi} MPC frameworks to name a few.

\textcolor{black}{In line with these developments, we adopt the viewpoint that safety-critical MPC can benefit from enforcing different state-constraint sets at different prediction steps. In CBF-based MPC, collision avoidance, and robotic receding-horizon planning, near-term predictions are closest to implementation and should therefore satisfy a certified safe set. Later predictions, however, can be constrained to a larger operating domain, since they will be re-optimized and re-certified before being implemented. This avoids enforcing tight safe set constraints over the full horizon, reducing conservativeness and computation while retaining near-term safety certification. Motivated by this perspective, and considering CBF constraints as invariant constraints along the MPC horizon,} our earlier work \cite{do2025constraint} introduced suboptimality estimation for ``partially constrained'' MPC, \textcolor{black}{where} a single invariant state-constraint set is enforced only over the constraint horizon, while the remaining predicted states are unconstrained with respect to that set. Relaxing \textcolor{black}{full-horizon constraint enforcement in \cite{do2025constraint}} yields results similar to \cite{grune2008infinite, grune2009practical}, \textcolor{black}{producing} a comparable \textcolor{black}{horizon-dependent} suboptimality bound. However, \textcolor{black}{focusing} on the constraint horizon and its effect on suboptimality and stability offers a new perspective on the topic.

\textcolor{black}{This work makes three contributions to} the suboptimality estimation landscape. First, we provide a generalized MPC formulation for discrete-time nonlinear systems where multiple state constraints are applied over the horizon, conceptually contributing to MPC frameworks \textcolor{black}{without terminal penalty or dedicated stabilizing terminal set, aligned with} \cite{limon2006stability, grune2017nonlinear,boccia2014stability}. \textcolor{black}{Distinct from} \cite{do2025constraint}, \textcolor{black}{which enforces one invariant set only over the constraint horizon, the present formulation allows a second, possibly less restrictive, state-constraint set over the remaining horizon. When this second set is chosen as a sufficiently large compact operating domain, the proposed formulation recovers, in set-enforcement terms and over the common horizon range, the partially constrained formulation
in~\cite{do2025constraint}}. Second, \textcolor{black}{in contrast to} \cite{grune2008infinite,grune2009practical, grune2018dynamic,worthmann2011stability,do2025constraint, worthmann2014role}, \textcolor{black}{our derivation exploits cancellations between shifted optimal/admissible costs, leading to a value-function-difference analysis rather than the induction-based mechanism used there to obtain suboptimality upper-bounds. This differentiates the roles of constraint-set selection and prediction-horizon length, gives qualitative insight into how they affect our suboptimality estimation, and produces implicit and horizon-explicit certificates. Although the main results developed here are stated uniformly over the invariant set, they can also be implemented in an \emph{a posteriori} trajectory-based fashion, being closer in spirit to} \cite{grune2009practical}. Compared to \cite{do2025constraint}, our method gives tighter bounds with a broader validity region \textcolor{black}{in the reported cases}. Third, we provide a lower bound expression on the closed-loop cost that can be tighter \textcolor{black}{under verifiable conditions} and is never weaker than the finite-horizon open-loop cost.

This paper is structured as follows: Section \ref{sec:formulation} formulates the problem. Section \ref{rdp_results} presents the Relaxed Dynamic programming (RDP) \cite{grune2008infinite} results \textcolor{black}{used in} Sections \ref{sec:cl_ubound} and \ref{sec:cl_lbound}, where closed-loop upper and lower bounds \textcolor{black}{are derived}, respectively. Section \ref{sec:simulations} presents linear and nonlinear safety-critical \textcolor{black}{simulations}. Section \ref{sec:conclusion} \textcolor{black}{draws conclusions on this work. Unless stated otherwise, the proof of all results can be found in the Appendix, following their order of appearance.}

\section{Problem Formulation}\label{sec:formulation}

Consider the discrete-time dynamical system 
\begin{equation}\label{eq:system}
x_{k+1}=f(x_k,u_k),
\end{equation}
where $x_k \in \textcolor{black}{\mathbb{R}^{n_x}}$, $u_k \in \mathcal{U}\subset\textcolor{black}{\mathbb{R}^{n_u}}$, with $\mathcal{U}$ being a compact set with non-empty interior and \textcolor{black}{\(0\in\mathcal U\)}\textcolor{black}{. Furthermore, $f:\mathbb{R}^{n_x} \times \mathcal{U} \rightarrow \mathbb{R}^{n_x}$ satisfies $f(0,0)=0$}. We consider an MPC problem at time $k$ with state $x_k$, where state constraints over the horizon $N$ are split in two: the first \textcolor{black}{$N-\tilde N+1$} predicted states are constrained by the \emph{control-invariant set} $\mathcal{X}_1$, and the last \textcolor{black}{$\tilde N-1$} predicted states are constrained by a \textcolor{black}{larger, not necessarily invariant,} set $\mathcal{X}_2$, with $0 \in \mathcal{X}_1 \subset \mathcal{X}_2 \subseteq \mathcal{D}$. \textcolor{black}{The set $\mathcal{D}\subset\textcolor{black}{\mathbb{R}^{n_x}}$ denotes the compact state domain with non-empty interior, within which the finite-horizon state constraints are specified, whereas $\mathcal{X}_2$ is the tail state constraint set.} \textcolor{black}{The choice $\mathcal{X}_2=\mathcal{D}$ corresponds to the case in which
the tail is not subject to any additional state constraint beyond
membership in this compact state domain. When this domain constraint is inactive, this recovers, in set-enforcement terms, the partially
constrained formulation of \cite{do2025constraint}.}
\textcolor{black}{We propose the} heterogeneously constrained MPC (HC-MPC) problem, formulated as
\begin{subequations}\label{eq:VN_Ntilde}
    \begin{align}
    &V_{N}^{\tilde N}(x_k):=\min_{u_h(n|x_k)}\quad \label{cost} \sum_{n=0}^{N-1}l(x_h(n|x_k),u_h(n|x_k)) \\
    &\mathrm{subject~to:} \nonumber\\
    & \label{dyn_constr} x_h(n+1|x_k)=f(x_h(n|x_k),u_h(n|x_k)),n=0,\ldots,N-1, \\
    \label{init_state}
    &x_h(0|x_k)=x_k,  \\
    \label{inp_constr}
    &u_h(n|x_k)\in\mathcal{U},n=0,\ldots,N-1, \\
    \label{cbf_constr}
    & x_h(n+1|x_k) \in \mathcal{X}_1, n=0,\ldots,N-\tilde N, \\
    \label{ctr2}
    & x_h(n+1|x_k) \in \mathcal{X}_2, n=N-\tilde N+1,\ldots,N-1.
\end{align}
\end{subequations}
We also introduce an auxiliary formulation, the uniformly constrained MPC (UC-MPC), which will be used for analysis purposes and bound derivations at later sections. UC-MPC is \emph{not} used to compute the control input, the online control is \emph{always} obtained by solving Problem \eqref{eq:VN_Ntilde}.
\begin{subequations}\label{eq:VN}
    \begin{align}
    &V_N(x_k):=\min_{u_{d}(n|x_k)}\quad \sum_{n=0}^{N-1}l(x_{d}(n|x_k),u_{d}(n|x_k)) \\
    &\mathrm{subject~to:} \nonumber\\
    & x_d(n+1|x_k)=f(x_d(n|x_k),u_{d}(n|x_k)),n=0,\ldots,N-1, \\
    &x_d(0|x_k)=x_k,  \\
    &u_d(n|x_k)\in\mathcal{U},n=0,\ldots,N-1, \\
    & x_d(n+1|x_k) \in \mathcal{X}_2, n=0,\ldots,N-1.
\end{align}
\end{subequations}
We will consider Problem \eqref{eq:VN_Ntilde} with $N \geq 2, 2 \leq \tilde{N} \leq N$ and Problem  \eqref{eq:VN}  with $N\geq1$. Note that \eqref{eq:VN} cannot be derived from \eqref{eq:VN_Ntilde}, as it requires $\tilde{N}=N+1$ in \eqref{eq:VN_Ntilde}, which is invalid. Thus, $\tilde{N}$ does not appear in the notation of \eqref{eq:VN}. We assume $x_k \in \mathcal{X}_1$, but we do not require any predicted state of \eqref{eq:VN} to belong to $\mathcal{X}_1$. \textcolor{black}{Here, $x_k$ denotes the closed-loop measured state of System \eqref{eq:system} at time $k$, while $x_h(n|x_k)$ in \eqref{eq:VN_Ntilde} and $x_d(n|x_k)$ in \eqref{eq:VN} denote open-loop states predicted $n$ steps ahead by the corresponding finite-horizon problems initialized at $x_k$. We emphasize that the notation ``$(n|x_k)$'' in the open-loop problems identifies $x_k$ as the initial condition of the finite-horizon problem, rather than implying that the open-loop variables are functions of $x_k$ alone.} Unless stated otherwise, we assume the following throughout:
\begin{assumption}[Positive definiteness of cost]\label{pdc}
The cost $l(x,u)$ is assumed to be continuous and positive definite for all $x \in \mathcal{D}$ and for all $u \in \mathcal{U}$, jointly with respect to both arguments\textcolor{black}{. This means that $l(\cdot,\cdot)=0$ if and only if $x=0$ and $u=0$. For any other argument pair $l(\cdot,\cdot)>0$.}  
\end{assumption}
\textcolor{black}{
\begin{assumption}[Well-posedness]\label{wellpd}
The set $\mathcal X_1$ is control invariant, i.e., for every $x\in\mathcal X_1$ there exists $u\in\mathcal U$ such that
$f(x,u)\in\mathcal X_1$. Moreover, the finite-horizon problems \eqref{eq:VN_Ntilde} and \eqref{eq:VN}, when initialized from states in
$\mathcal X_1$, attain their minima for all horizons considered.
\end{assumption}
Assumption \ref{wellpd} is slightly stronger than the traditional viability assumption \cite{grune2017nonlinear,grune2008infinite,lorenzen2019stochastic}, as we require invariance of $\mathcal{X}_1$. Many technicalities simplify based on this stronger Assumption. Since $\mathcal X_1\subset\mathcal X_2$, the invariance of $\mathcal X_1$ provides an admissible input sequence that keeps the predicted trajectory in $\mathcal X_1$  (not necessarily the optimal one), and thus also satisfies the $\mathcal X_2$ constraints.
Thus, feasibility of both \eqref{eq:VN_Ntilde} and
\eqref{eq:VN} follow for initial conditions in $\mathcal X_1$. Existence of minimizers for these finite-horizon problems is handled by the attainment requirement in Assumption \ref{wellpd}. This setting allows us to produce a general formulation that complements existing MPC-CBF strategies \cite{zeng2021safety}, and increases the analytical and implementation flexibility of predictive safety filters \cite{wabersich2021predictive}.}

Let $[u^*_h(0|x_k),\ldots,u^*_h(N-1|x_k)]$ be the \emph{open-loop optimal control sequence}, generating the \emph{open-loop optimal trajectory} $[x^*_h(1|x_k),\ldots,x^*_h(N|x_k)]$ for problem \eqref{eq:VN_Ntilde}. For future analysis, we introduce their respective counterparts $[u^*_d(0|x_k),\ldots,u^*_d(N-1|x_k)]$ and $[x^*_d(1|x_k),\ldots,x^*_d(N|x_k)]$ obtained from \eqref{eq:VN}. Let
\begin{equation}
\mu^{\tilde{N}}_{N}(x_k):=u_h^*(0|x_k), \label{eq:closed-loop-control} 
\end{equation}
be the \emph{closed-loop} controller, defined to be $\mu^{\tilde{N}}_{N}(x_k)$ and \emph{always} obtained by solving \eqref{eq:VN_Ntilde}. This controller is applied to system \eqref{eq:system}, producing the closed-loop system dynamics
\begin{equation}\label{eq:closed-loop-dynamics}
    x_{k+1} = f(x_k,\mu^{\tilde{N}}_{N}(x_k)).
\end{equation}
Figure \ref{fig:OLscheme} shows the system behaviour under \eqref{eq:VN_Ntilde}. 

\begin{figure}[!tbp]
    \centering
    \resizebox{\columnwidth}{!}{%
        \begin{tikzpicture}[x=0.42pt,y=0.42pt,yscale=-1,xscale=1]

\draw  [color={rgb, 255:red, 0; green, 0; blue, 0 }  ,draw opacity=1 ][fill={rgb, 255:red, 180; green, 54; blue, 48 }  ,fill opacity=1 ][line width=2.25]  (158.5,264.21) .. controls (158.5,210.91) and (201.7,167.71) .. (255,167.71) -- (817.59,167.71) .. controls (870.89,167.71) and (914.09,210.91) .. (914.09,264.21) -- (914.09,553.7) .. controls (914.09,607) and (870.89,650.2) .. (817.59,650.2) -- (255,650.2) .. controls (201.7,650.2) and (158.5,607) .. (158.5,553.7) -- cycle ;
\draw  [color={rgb, 255:red, 52; green, 98; blue, 152 }  ,draw opacity=1 ][fill={rgb, 255:red, 74; green, 135; blue, 206 }  ,fill opacity=1 ][line width=2.25]  (357.2,245.8) .. controls (423.2,231.8) and (449.2,290.8) .. (530.2,221.8) .. controls (611.2,152.8) and (664.2,157.8) .. (697.2,265.8) .. controls (730.2,373.8) and (877.2,488.8) .. (751.2,561.8) .. controls (625.2,634.8) and (329.2,652.8) .. (276,565.2) .. controls (222.8,477.6) and (291.2,259.8) .. (357.2,245.8) -- cycle ;
\draw  [color={rgb, 255:red, 65; green, 117; blue, 5 }  ,draw opacity=1 ][fill={rgb, 255:red, 150; green, 200; blue, 97 }  ,fill opacity=1 ][line width=2.25]  (338.89,515.9) .. controls (309.24,469.7) and (354.09,388.03) .. (439.07,333.5) .. controls (524.04,278.97) and (616.97,272.22) .. (646.62,318.42) .. controls (676.27,364.63) and (631.42,446.29) .. (546.44,500.82) .. controls (461.47,555.36) and (368.54,562.11) .. (338.89,515.9) -- cycle ;
\draw  [dash pattern={on 2.53pt off 3.02pt}][line width=2.25]  (384.2,441.2) .. controls (384.2,437.89) and (386.89,435.2) .. (390.2,435.2) .. controls (393.51,435.2) and (396.2,437.89) .. (396.2,441.2) .. controls (396.2,444.51) and (393.51,447.2) .. (390.2,447.2) .. controls (386.89,447.2) and (384.2,444.51) .. (384.2,441.2) -- cycle ;
\draw [line width=2.25]    (390.2,441.2) -- (526.2,452.6) ;
\draw [shift={(526.2,452.6)}, rotate = 323.13] [color={rgb, 255:red, 0; green, 0; blue, 0 }  ][fill={rgb, 255:red, 0; green, 0; blue, 0 }  ][line width=2.25]      (0, 0) circle [x radius= 5.36, y radius= 5.36]   ;
\draw [line width=2.25]    (526.2,452.6) -- (585.2,345.6) ;
\draw [shift={(585.2,345.6)}, rotate = 190.52] [color={rgb, 255:red, 0; green, 0; blue, 0 }  ][fill={rgb, 255:red, 0; green, 0; blue, 0 }  ][line width=2.25]      (0, 0) circle [x radius= 5.36, y radius= 5.36]   ;
\draw [line width=2.25]    (585.2,345.6) -- (468.2,374.6) ;
\draw [shift={(468.2,374.6)}, rotate = 219.67] [color={rgb, 255:red, 0; green, 0; blue, 0 }  ][fill={rgb, 255:red, 0; green, 0; blue, 0 }  ][line width=2.25]      (0, 0) circle [x radius= 5.36, y radius= 5.36]   ;
\draw [line width=2.25]    (468.2,374.6) -- (522.2,278.6) ;
\draw [shift={(522.2,278.6)}, rotate = 1.47] [color={rgb, 255:red, 0; green, 0; blue, 0 }  ][fill={rgb, 255:red, 0; green, 0; blue, 0 }  ][line width=2.25]      (0, 0) circle [x radius= 5.36, y radius= 5.36]   ;
\draw [line width=2.25]    (522.2,278.6) -- (589.00,211.76) ;
\draw [shift={(589.00,211.76)}, rotate = 17.31] [color={rgb, 255:red, 0; green, 0; blue, 0 }  ][fill={rgb, 255:red, 0; green, 0; blue, 0 }  ][line width=2.25]      (0, 0) circle [x radius= 5.36, y radius= 5.36]   ;
\draw [line width=2.25]    (589.00,211.76) -- (678.2,381.6) ;
\draw [shift={(678.2,381.6)}, rotate = 121.61] [color={rgb, 255:red, 0; green, 0; blue, 0 }  ][fill={rgb, 255:red, 0; green, 0; blue, 0 }  ][line width=2.25]      (0, 0) circle [x radius= 5.36, y radius= 5.36]   ;

\draw (199.84,209.52) node [anchor=north west][inner sep=0.75pt]  [font=\normalsize,color={rgb, 255:red, 255; green, 255; blue, 255 }  ,opacity=1 ]  {$\mathcal{D}$};
\draw (357.29,504.36) node [anchor=north west][inner sep=0.75pt]  [font=\normalsize]  {$\mathcal{X}_{1}$};
\draw (338.29,285.36) node [anchor=north west][inner sep=0.75pt]  [font=\normalsize]  {$\mathcal{X}_{2}$};
\draw (379,444.4) node [anchor=north west][inner sep=0.75pt]  [font=\normalsize]  {$x_{k}$};
\draw (468.48,493.06) node [anchor=north west][inner sep=0.75pt]  [font=\normalsize,rotate=-327.68]  {$x_{k+1} =x_{h}( 1|x_{k})$};
\draw (535.88,310.14) node [anchor=north west][inner sep=0.75pt]  [font=\normalsize,rotate=-359.35]  {$x_{h}( 2|x_{k})$};
\draw (431.88,389.14) node [anchor=north west][inner sep=0.75pt]  [font=\normalsize,rotate=-359.35]  {$x_{h}( 3|x_{k})$};
\draw (455.06,269.17) node [anchor=north west][inner sep=0.75pt]  [font=\normalsize,rotate=-332.03]  {$x_{h}( 4|x_{k})$};
\draw (613.2,205.6) node [anchor=north west][inner sep=0.75pt]  [font=\normalsize,rotate=-32.00]  {$x_{h}( 5|x_{k})$};
\draw (647.82,401.93) node [anchor=north west][inner sep=0.75pt]  [font=\normalsize,rotate=-351.56]  {$x_{h}( 6|x_{k})$};
\end{tikzpicture}
    }
    \caption{System \eqref{eq:system} under MPC \eqref{eq:VN_Ntilde} for $N=6$, $N-\tilde{N}=2$. Input $u_h(2|x_k)$ produces the last open loop state $x_h(3|x_k)$ in $\mathcal{X}_1$. Subsequent open loop states are subject to $\mathcal{X}_2$. Closed-loop states are depicted by $x_k$ and $x_{k+1}$. }
    \label{fig:OLscheme}
\end{figure}


The associated \emph{closed-loop infinite horizon cost} of \eqref{eq:VN_Ntilde} is defined by
\begin{equation}\label{eq:closed-loop-cost}
    J_{\infty}^{N,\tilde N}(x_0)=\sum_{k=0}^\infty l(x_k,\mu^{\tilde{N}}_N(x_k)).
\end{equation}
Our goal is to investigate upper \cite{grune2008infinite} and lower bounds for $J_\infty^{N,\tilde N}(x_0)$, \textcolor{black}{which is hard to obtain \emph{a priori}}, and to explore the impact of the heterogeneous constraint scheme presented. Beyond the tighter upper-bound results and theoretical 
novelty of the lower bound per se, obtaining both bounds works as a powerful design tool. We start our investigation by deriving bounds using Relaxed Dynamic Programming (RDP). 

\section{Closed-loop bounds based on Relaxed Dynamic programming} \label{rdp_results}
\subsection{Closed-loop upper-bound}
\textcolor{black}{We start by  stating the following result, whose proof can be found in~\cite[Proposition 2.2]{grune2008infinite}, \cite{do2025constraint}.}  
\begin{lemma}\label{lem:cl-ub-optimality}
    Consider $N \geq 2, 2 \leq \tilde{N} \leq N$. Let the following Relaxed Dynamic Programming (RDP) equation
\begin{equation} \label{eq:dyn_prog}
        V_N^{\tilde N}(x_k)\ge V_N^{\tilde N}(x_{k+1})+\alpha l(x_k,\mu^{\tilde{N}}_{N}(x_k)),
    \end{equation}
hold for some $\alpha \in [0,1]$ and for all $x_k \in \mathcal{X}_1$. Then 
    \begin{equation}\label{eq:alpha}
        \alpha J_{\infty}^{N,\tilde N}(x_k)\le V_N^{\tilde N}(x_k),
    \end{equation}
    holds for all $x_k\in\mathcal{X}_1$.
\end{lemma}
\textcolor{black}{This result links closed-loop information in \eqref{eq:closed-loop-cost} to open-loop information via the upper bound \eqref{eq:alpha}.  The case $\alpha=0$ is retained here only as a boundary case of the RDP inequality. It states non-increasing $V_N^{\tilde N}(\cdot)$, but no meaningful closed-loop cost bound is obtained in this case. Therefore, a meaningful suboptimality bound certificate requires $\alpha>0$.}

\subsection{Closed-loop lower bound}
For the lower bound on \eqref{eq:closed-loop-cost}, a natural first choice is
\(V^{2}_{\infty}(x_k)\), \textcolor{black}{where $V^{\tilde{N}}_{\infty}(x_k)\coloneqq \lim_{M \to \infty} V^{\tilde{N}}_M(x_k)$, with fixed $\tilde{N}$, } mirroring the standard lower bound
\(V_{\infty}(x_k)\) for the traditional formulation; see,
e.g., \cite[Prop.~2.2]{grune2008infinite},
\cite[Lemma~4]{grune2009practical}. However, since $V^{2}_{\infty}(x_k)$ can only be approximated by solving \eqref{eq:VN_Ntilde} with a very large horizon $N \gg \tilde{N}$, its computation is often expensive. A looser but computationally cheaper lower bound is, $V^{\tilde{N}}_N (x_k)$, which holds via \begin{equation}\label{lb_sec}
    J^{N,\tilde{N}}_{\infty}(x_k) \geq V^{2}_{\infty}(x_k) \geq V^{\tilde{N}}_{\infty}(x_k) \geq V^{\tilde{N}}_{N}(x_k),
\end{equation}
and is valid as our problem formulations do not use a terminal cost \cite{grune2008infinite}.  
We seek a higher lower bound on \eqref{eq:closed-loop-cost} than $V_N^{\tilde N}(x_k)$. 

\textcolor{black}{In the traditional upper-bound argument (e.g.: \eqref{eq:alpha}), the nonnegative terminal value
$V_N^{\tilde N}(x_{k+1+M})$ can be discarded at the finite-$M$
telescoping step before taking the limit $M\to\infty$ \cite{grune2009analysis}. For the lower bound results however, the same approach cannot be used, as we need it to vanish as $M\to\infty$. The following lemma, based on the standard Lyapunov/RDP
convergence argument used in nonlinear MPC, provides this property.}

\begin{lemma} \label{lemma_2}
Let $N \geq 2$, $2 \leq \tilde{N} \leq N$, and suppose that, for all 
$x_k \in \mathcal{X}_1$, \eqref{eq:dyn_prog} holds for some 
$\textcolor{black}{\alpha \in (0,1]}$.  
\textcolor{black}{Assume also that $V_N^{\tilde N}(\cdot)$ is continuous at the origin.} Then, \textcolor{black}{for every closed-loop trajectory
initialized in $\mathcal{X}_1$}, if $M \rightarrow \infty$, 
$V_N^{\tilde N}(x_{k+1+M}) \rightarrow 0$.
\end{lemma}
The restriction to MPC problems with positive definite costs \textcolor{black}{(as per Assumption \ref{pdc})} is essential here. Indeed, by 
\eqref{eq:dyn_prog}, 
\begin{equation}\label{eq:Lyap}
V_N^{\tilde N}(x_{k+1}) - V_N^{\tilde N}(x_k)
    \leq -\alpha l(x_k,\mu^{\tilde{N}}_{N}(x_k)).
\end{equation}    
\textcolor{black}{For the proof see \cite[Theorem~4.11]{grne2013nonlinear}. Since $\alpha>0$ and $l(\cdot,\cdot)$ is positive definite, \eqref{eq:Lyap} acts as a discrete time Lyapunov equation. Thus, the standard Lyapunov/RDP argument implies convergence of the closed-loop state to the origin. The additional continuity condition on $V_N^{\tilde N}(x_k)$ at the origin then yields $V_N^{\tilde N}(x_{k+1+M})\rightarrow0$ as $M\rightarrow\infty$. Consequently, $\alpha>0$ guarantees asymptotic stability, while $\alpha=0$ yields Lyapunov stability, but produces a degenerate performance bound.} Via RDP, we derive results analogous to the upper bound \eqref{eq:alpha}. 
\begin{proposition} \label{prop_lb}
Let $N \geq 2$, $2 \leq \tilde{N} \leq N$
\textcolor{black}{, and assume $V^{\tilde N}_N(\cdot)$ to be continuous at the origin.}
\textcolor{black}{Suppose that \eqref{eq:dyn_prog} holds for some $\alpha\in(0,1]$ and for all $x_k\in\mathcal X_1$.} 
Let the following \textcolor{black}{lower} relaxed dynamic programming inequality \textcolor{black}{also} hold for
\textcolor{black}{some $\omega \in [0,1)$ and for all $x_k\in\mathcal X_1$}
\begin{subequations} 
\begin{gather} \label{eq:dyn_prog_2}
\textcolor{black}{
l(x_k,\mu^{\tilde{N}}_{N}(x_k)) \geq
\frac{1}{1-\omega} \left[
V_N^{\tilde N}(x_k)- V_N^{\tilde N}(x_{k+1})
\right].
}
\end{gather}
\textrm{Then: \nonumber}
\begin{gather}
\alpha J_{\infty}^{N,\tilde N}(x_k) \geq
\frac{\alpha}{1-\omega}V_N^{\tilde N}(x_k). \label{eq:partial_lb}
\end{gather}
\textcolor{black}{Moreover, since both inequalities refer to the same $V_N^{\tilde N}(\cdot)$, $\mu_N^{\tilde N}(\cdot)$, and horizon pair $(N,\tilde N)$, \eqref{eq:partial_lb} combines directly with \eqref{eq:dyn_prog}, yielding}
\begin{gather}
\frac{V_N^{\tilde N}(x_k)}{\alpha} \geq  J_{\infty}^{N,\tilde N}(x_k) \geq
\frac{1}{1-\omega}V_N^{\tilde N}(x_k), \label{eq:clbd}
\end{gather}
\textrm{for all $x_k\in\mathcal{X}_1$}. \nonumber
\end{subequations} 
\end{proposition}

\textcolor{black}{Hence, for a fixed horizon pair $(N,\tilde N)$, whenever \eqref{eq:dyn_prog} holds with some $\alpha\in(0,1]$ and \eqref{eq:dyn_prog_2} holds with some $\omega\in[0,1)$ for the same value function
$V_N^{\tilde N}(x_k)$ and feedback law $\mu_N^{\tilde N}(x_k)$, the two-sided estimate
\eqref{eq:clbd} holds with no extra compatibility condition between $\alpha$ and $\omega$  required.}
Sections \ref{sec:cl_ubound} and \ref{sec:cl_lbound} characterize $\alpha$ and $\omega$, respectively: First via direct solution of \eqref{eq:VN_Ntilde} and \eqref{eq:VN} and, under extra cost controllability assumptions, explicitly in terms of the pair $(N,\tilde{N})$.

\section{Closed-loop performance upper bound}
\label{sec:cl_ubound}

\subsection{Calculation of $\alpha$ as a difference of value functions}
We want to express $\alpha$ as a function of $(N,\tilde N)$, which will happen at first implicitly, based on a difference of value functions. This requires the analysis of $V_N^{\tilde N}(x_k)-V_N^{\tilde N}(x_{k+1})$, which we do by defining the optimal open-loop state–input pair of \eqref{eq:VN_Ntilde} and \eqref{eq:VN}:
\textcolor{black}{
\begin{subequations}
    \begin{align}
    & V_N^{\tilde N}(x_k) = \sum^{N-\tilde{N}}_{n=0} l(x^*_h(n|x_k),u^*_h(n|x_k)) \nonumber \\
    & + \sum^{N-1}_{n=N-\tilde{N}+1} l(x^*_h(n|x_k),u^*_h(n|x_k)).  \label{eq:brkdwn} \\ 
    & V_N(x_k) = \sum^{N-1}_{n=0} l(x^*_d(n|x_k),u^*_d(n|x_k)). \label{eq:opt_seq}
    \end{align}
\end{subequations}
}
\textcolor{black}{
In \eqref{eq:brkdwn} we split the sum to highlight a specific characteristic which will be important in our derivations. The first summation contains the running costs using inputs $u^*_h(n|x_k)$ that are able to produce $x^*_h(n+1|x_k)\in \mathcal{X}_1$, whereas the second summation uses $u^*_h(n|x_k)$ enforcing $x^*_h(n+1|x_k)\in \mathcal{X}_2$. In \eqref{eq:opt_seq} we do not perform the same split as $u^*_d(n|x_k)$ always generates states in $\mathcal{X}_2$. From now on, we will use 
\begin{equation*}
   \lambda_n \coloneqq l(x^*_{h}(n|x_k),u^*_h(n|x_k)), n=0,\dots, N-\tilde{N}, 
\end{equation*}
as a short hand in our derivations.} To derive an expression for $\alpha$, the following Lemmas will allow us to bound $V_N^{\tilde N}(x_k)-V_N^{\tilde N}(x_{k+1})$ with a difference between the value functions of problems \eqref{eq:VN_Ntilde} and \eqref{eq:VN}.
\begin{lemma}\label{lem:eqVNtilde}
    \textcolor{black}{Let Assumption \ref{wellpd} hold.} Then, $\textcolor{black}{\sum^{N-1}_{n=N-\tilde{N}+1} l(x^*_h(n|x_k),u^*_h(n|x_k))} = V_{\tilde{N}-1}(x^*_{h}(N-\tilde{N}+1|x_k))$. 
\end{lemma}
\begin{lemma}\label{lem:diff_val_func_ubd} \textcolor{black}{Let Assumption \ref{wellpd} hold.} For $N \geq 3, 2 \leq \tilde{N} \leq N-1$, we can lower bound $V_N^{\tilde N}(x_k) - V_N^{\tilde N}(x_{k+1})$ as:
    \begin{align}\label{eq:lem4}
        & V_N^{\tilde N}(x_k) - V_N^{\tilde N}(\textcolor{black}{x_{k+1})\ge \lambda_0} \nonumber \\
        & -\left (V_{\tilde{N}}^{\tilde{N}}(x^*_{h}(N-\tilde{N}+1|x_k)) - V_{\tilde{N}-1}(x^*_{h}(N-\tilde{N}+1|x_k)) \right).
    \end{align}
\end{lemma}
\textcolor{black}{The horizon restrictions of Lemma \ref{lem:diff_val_func_ubd}, $N \geq 3, 2 \leq \tilde{N} \leq N-1$, are deliberately retained whenever its bound \eqref{eq:lem4} is used in subsequent results. If $\tilde N=N$, the shifted-trajectory argument remains valid, but $x_h^*(N-\tilde N+1\mid x_k)=x_{k+1}$ and the bound in \eqref{eq:lem4} becomes an equality. Thus, obtaining an admissible $\alpha$ from \eqref{eq:lem4} is equivalent to verifying \eqref{eq:dyn_prog} directly. Thus, this boundary case is not pursued in those results.} Using Lemmas \ref{lem:eqVNtilde}, \ref{lem:diff_val_func_ubd}, \textcolor{black}{and aiming to establish the RDP inequality \eqref{eq:dyn_prog}, it is sufficient to ensure the following inequality chain:
    \begin{align} \label{eq:upbound_dif}
        & V_N^{\tilde N}(x_k) - V_N^{\tilde N}(x_{k+1})\ge \lambda_0 \nonumber \\
        & -\left (V_{\tilde{N}}^{\tilde{N}}(x^*_{h}(N-\tilde{N}+1|x_k)) - V_{\tilde{N}-1}(x^*_{h}(N-\tilde{N}+1|x_k)) \right) \nonumber \\
        & \geq \alpha \lambda_0.
    \end{align}
Since $\lambda_0 = l(x_k,\mu_N^{\tilde N}(x_k))$, whenever the intermediate inequality in \eqref{eq:upbound_dif} is nonnegative, \eqref{eq:upbound_dif} yields an admissible $\alpha\in[0,1]$ and
therefore implies \eqref{eq:dyn_prog}. With that, we can state our first $\alpha$ estimate.
}
\textcolor{black}{
\begin{lemma}\label{lem:implicit_alpha}
For $N \geq 3$, $2 \leq \tilde N \leq N-1$, suppose that, for all 
$x_k\in\mathcal X_1$, the intermediate lower bound in 
\eqref{eq:upbound_dif} is nonnegative. Then, a uniform RDP 
certificate can be obtained by
\begin{align} \label{opt_alpha}
     \alpha & = 1 - \sup_{x_k\in\mathcal X_1\setminus\{0\}}\frac{1}{\lambda_0}
     \left [ V_{\tilde{N}}^{\tilde{N}}(x^*_{h}(N-\tilde{N}+1|x_k)) \right. \nonumber \\
    & \left. - V_{\tilde{N}-1}(x^*_{h}(N-\tilde{N}+1|x_k)) \right].
\end{align}
If $\alpha\in[0,1]$, then \eqref{eq:dyn_prog} holds for all 
$x_k\in\mathcal X_1$, and the conditions of Lemma 
\ref{lem:cl-ub-optimality} are fulfilled.
\end{lemma}
}
\textcolor{black}{Lemma~\ref{lem:implicit_alpha} provides an implicit quantitative construction of a uniform certificate $\alpha$. The resulting suboptimality level 
is determined by the difference between the value functions of 
Problem~\eqref{eq:VN_Ntilde} with horizon $\tilde N$, and Problem~\eqref{eq:VN} with horizon $\tilde N-1$.} \textcolor{black}{In a set-wise verification over \(\mathcal X_1\), this avoids the direct
evaluation of \(V_N^{\tilde N}(x_{k+1})\) for each sampled \(x_k\), replacing it with auxiliary value-function evaluations of shorter horizons
\(\tilde N\) and \(\tilde N-1\) at the switching state \(x_s\).} A visual explanation of Lemma \ref{lem:implicit_alpha} is shown in Figure \ref{fig:canIdea}.

\begin{figure}[htbp]
    \centering
    \resizebox{\columnwidth}{!}{%
   
\tikzset{every picture/.style={font=\LARGE}} 

\begin{tikzpicture}[x=0.75pt,y=0.75pt,yscale=-1,xscale=1]

\draw [line width=1.5]  [dash pattern={on 5.63pt off 4.5pt}]  (230.6,288) -- (341.8,288) ;
\draw [shift={(345.8,288)}, rotate = 180] [fill={rgb, 255:red, 0; green, 0; blue, 0 }  ][line width=0.08]  [draw opacity=0] (11.61,-5.58) -- (0,0) -- (11.61,5.58) -- cycle    ;
\draw [shift={(230.6,288)}, rotate = 0] [color={rgb, 255:red, 0; green, 0; blue, 0 }  ][fill={rgb, 255:red, 0; green, 0; blue, 0 }  ][line width=1.5]      (0, 0) circle [x radius= 4.36, y radius= 4.36]   ;
\draw [line width=1.5]    (652.2,287) -- (763.4,287) ;
\draw [shift={(767.4,287)}, rotate = 180] [fill={rgb, 255:red, 0; green, 0; blue, 0 }  ][line width=0.08]  [draw opacity=0] (11.61,-5.58) -- (0,0) -- (11.61,5.58) -- cycle    ;
\draw [shift={(652.2,287)}, rotate = 0] [color={rgb, 255:red, 0; green, 0; blue, 0 }  ][fill={rgb, 255:red, 0; green, 0; blue, 0 }  ][line width=1.5]      (0, 0) circle [x radius= 4.36, y radius= 4.36]   ;
\draw [line width=1.5]    (497,288) -- (608.2,288) ;
\draw [shift={(612.2,288)}, rotate = 180] [fill={rgb, 255:red, 0; green, 0; blue, 0 }  ][line width=0.08]  [draw opacity=0] (11.61,-5.58) -- (0,0) -- (11.61,5.58) -- cycle    ;
\draw [shift={(497,288)}, rotate = 0] [color={rgb, 255:red, 0; green, 0; blue, 0 }  ][fill={rgb, 255:red, 0; green, 0; blue, 0 }  ][line width=1.5]      (0, 0) circle [x radius= 4.36, y radius= 4.36]   ;
\draw [line width=2.25]  [dash pattern={on 2.53pt off 3.02pt}]  (497,228.7) -- (497,347.3) ;
\draw [line width=2.25]  [dash pattern={on 2.53pt off 3.02pt}]  (767.4,227.7) -- (767.4,346.3) ;
\draw [line width=1.5]  [dash pattern={on 5.63pt off 4.5pt}]  (381.8,288) -- (493,288) ;
\draw [shift={(497,288)}, rotate = 180] [fill={rgb, 255:red, 0; green, 0; blue, 0 }  ][line width=0.08]  [draw opacity=0] (11.61,-5.58) -- (0,0) -- (11.61,5.58) -- cycle    ;
\draw [shift={(381.8,288)}, rotate = 0] [color={rgb, 255:red, 0; green, 0; blue, 0 }  ][fill={rgb, 255:red, 0; green, 0; blue, 0 }  ][line width=1.5]      (0, 0) circle [x radius= 4.36, y radius= 4.36]   ;
\draw [line width=1.5]    (115.4,288) -- (226.6,288) ;
\draw [shift={(230.6,288)}, rotate = 180] [fill={rgb, 255:red, 0; green, 0; blue, 0 }  ][line width=0.08]  [draw opacity=0] (11.61,-5.58) -- (0,0) -- (11.61,5.58) -- cycle    ;
\draw [shift={(115.4,288)}, rotate = 0] [color={rgb, 255:red, 0; green, 0; blue, 0 }  ][fill={rgb, 255:red, 0; green, 0; blue, 0 }  ][line width=1.5]      (0, 0) circle [x radius= 4.36, y radius= 4.36]   ;
\draw [color={rgb, 255:red, 51; green, 26; blue, 180 }  ,draw opacity=1 ][line width=2.25]
    (230.6,288) -- (230.6,604) ;
\draw [shift={(230.6,604)}, rotate = 270] [fill={rgb, 255:red, 51; green, 26; blue, 180 }  ,fill opacity=1 ][line width=0.08]  [draw opacity=0]
    (14.29,-6.86) -- (0,0) -- (14.29,6.86) -- cycle ;
\draw [shift={(230.6,288)}, rotate = 90] [fill={rgb, 255:red, 51; green, 26; blue, 180 }  ,fill opacity=1 ][line width=0.08]  [draw opacity=0]
    (14.29,-6.86) -- (0,0) -- (14.29,6.86) -- cycle ;
\draw [color={rgb, 255:red, 174; green, 14; blue, 34 }  ,draw opacity=1 ][line width=1.5]    (499,606) -- (610.2,606) ;
\draw [shift={(614.2,606)}, rotate = 180] [fill={rgb, 255:red, 174; green, 14; blue, 34 }  ,fill opacity=1 ][line width=0.08]  [draw opacity=0] (11.61,-5.58) -- (0,0) -- (11.61,5.58) -- cycle    ;
\draw [shift={(499,606)}, rotate = 0] [color={rgb, 255:red, 174; green, 14; blue, 34 }  ,draw opacity=1 ][fill={rgb, 255:red, 174; green, 14; blue, 34 }  ,fill opacity=1 ][line width=1.5]      (0, 0) circle [x radius= 4.36, y radius= 4.36]   ;
\draw [line width=2.25]  [dash pattern={on 2.53pt off 3.02pt}]  (614.2,546.7) -- (614.2,665.3) ;
\draw [color={rgb, 255:red, 174; green, 14; blue, 34 }  ,draw opacity=1 ][line width=1.5]    (765.2,607) -- (805.2,607) -- (876.4,607) ;
\draw [shift={(880.4,607)}, rotate = 180] [fill={rgb, 255:red, 174; green, 14; blue, 34 }  ,fill opacity=1 ][line width=0.08]  [draw opacity=0] (11.61,-5.58) -- (0,0) -- (11.61,5.58) -- cycle    ;
\draw [shift={(765.2,607)}, rotate = 0] [color={rgb, 255:red, 174; green, 14; blue, 34 }  ,draw opacity=1 ][fill={rgb, 255:red, 174; green, 14; blue, 34 }  ,fill opacity=1 ][line width=1.5]      (0, 0) circle [x radius= 4.36, y radius= 4.36]   ;
\draw [line width=2.25]  [dash pattern={on 2.53pt off 3.02pt}]  (880.4,547.7) -- (880.4,666.3) ;
\draw  [line width=2.25]  (497,222.7) .. controls (572.11,213.75) and (617.18,204.81) .. (632.2,195.87) .. controls (647.23,204.81) and (692.29,213.75) .. (767.4,222.7) ;
\draw  [color={rgb, 255:red, 174; green, 14; blue, 34 }  ,draw opacity=1 ][line width=2.25]  (880.4,668.3) .. controls (774.46,679.88) and (710.89,691.46) .. (689.7,703.03) .. controls (668.51,691.46) and (604.95,679.88) .. (499,668.3) ;
\draw [color={rgb, 255:red, 174; green, 14; blue, 34 }  ,draw opacity=1 ][line width=2.25]  [dash pattern={on 6.75pt off 4.5pt}]  (499,606) -- (499,665.3) ;
\draw [color={rgb, 255:red, 174; green, 14; blue, 34 }  ,draw opacity=1 ][line width=1.5]  [dash pattern={on 5.63pt off 4.5pt}]  (230.6,604) -- (341.8,604) ;
\draw [shift={(345.8,604)}, rotate = 180] [fill={rgb, 255:red, 174; green, 14; blue, 34 }  ,fill opacity=1 ][line width=0.08]  [draw opacity=0] (11.61,-5.58) -- (0,0) -- (11.61,5.58) -- cycle    ;
\draw [shift={(230.6,604)}, rotate = 0] [color={rgb, 255:red, 174; green, 14; blue, 34 }  ,draw opacity=1 ][fill={rgb, 255:red, 174; green, 14; blue, 34 }  ,fill opacity=1 ][line width=1.5]      (0, 0) circle [x radius= 4.36, y radius= 4.36]   ;
\draw [color={rgb, 255:red, 174; green, 14; blue, 34 }  ,draw opacity=1 ][line width=1.5]  [dash pattern={on 5.63pt off 4.5pt}]  (381.8,606) -- (493,606) ;
\draw [shift={(497,606)}, rotate = 180] [fill={rgb, 255:red, 174; green, 14; blue, 34 }  ,fill opacity=1 ][line width=0.08]  [draw opacity=0] (11.61,-5.58) -- (0,0) -- (11.61,5.58) -- cycle    ;
\draw [shift={(381.8,606)}, rotate = 0] [color={rgb, 255:red, 174; green, 14; blue, 34 }  ,draw opacity=1 ][fill={rgb, 255:red, 174; green, 14; blue, 34 }  ,fill opacity=1 ][line width=1.5]      (0, 0) circle [x radius= 4.36, y radius= 4.36]   ;
\draw [color={rgb, 255:red, 174; green, 14; blue, 34 }  ,draw opacity=1 ][line width=1.5]    (651.2,607) -- (691.2,607) -- (762.4,607) ;
\draw [shift={(766.4,607)}, rotate = 180] [fill={rgb, 255:red, 174; green, 14; blue, 34 }  ,fill opacity=1 ][line width=0.08]  [draw opacity=0] (11.61,-5.58) -- (0,0) -- (11.61,5.58) -- cycle    ;
\draw [shift={(651.2,607)}, rotate = 0] [color={rgb, 255:red, 174; green, 14; blue, 34 }  ,draw opacity=1 ][fill={rgb, 255:red, 174; green, 14; blue, 34 }  ,fill opacity=1 ][line width=1.5]      (0, 0) circle [x radius= 4.36, y radius= 4.36]   ;

\draw (630,288) node [anchor=center][inner sep=0.75pt]  [font=\Large]  {$\dotsc $};
\draw (506,225.4) node [anchor=north west][inner sep=0.75pt]    {$\mathcal{X}_{1}$};
\draw (774,226.4) node [anchor=north west][inner sep=0.75pt]    {$\mathcal{X}_{2}$};
\draw (360,288) node [anchor=center][inner sep=0.75pt]  [font=\Large]  {$\dotsc $};
\draw (360,606) node [anchor=center][inner sep=0.75pt]  [font=\Large,color={rgb, 255:red, 174; green, 14; blue, 34 }  ,opacity=1 ]  {$\dotsc $};
\draw (620,544.4) node [anchor=north west][inner sep=0.75pt]    {$\mathcal{X}_{1}$};
\draw (630,606) node [anchor=center][inner sep=0.75pt]  [font=\Large,color={rgb, 255:red, 174; green, 14; blue, 34 }  ,opacity=1 ]  {$\dotsc $};
\draw (887,546.4) node [anchor=north west][inner sep=0.75pt]    {$\mathcal{X}_{2}$};
\draw (126.89,293.04) node [anchor=north west][inner sep=0.75pt]  [font=\Large,rotate=-70]  {$x_{k} =x^*_{h}\left( 0|x_{k}\right)$};
\draw (255.09,293.04) node [anchor=north west][inner sep=0.75pt]  [font=\Large,rotate=-70]  {$x_{k+1} =x^*_{h}\left( 1|x_{k}\right)$};
\draw (404.29,293.04) node [anchor=north west][inner sep=0.75pt]  [font=\Large,rotate=-70]  {$x^*_{h}\left( N-\tilde{N} |x_{k}\right)$};
\draw (525.49,293.04) node [anchor=north west][inner sep=0.75pt]  [font=\Large,rotate=-62]  {$x^*_{h}\left( N-\tilde{N} +1|x_{k}\right)$};
\draw (669.49,295.04) node [anchor=north west][inner sep=0.75pt]  [font=\Large,rotate=-70]  {$x^*_{h}( N-1|x_{k})$};
\draw (789.49,295.04) node [anchor=north west][inner sep=0.75pt]  [font=\Large,rotate=-70]  {$x^*_{h}( N|x_{k})$};
\draw (159.97,267.17) node [anchor=north west][inner sep=0.75pt]  [font=\Large,rotate=-290]  {$l\left( x_{h}^{*}( 0|x_{k}) ,u_h^*( 0|x_{k})\right)$};
\draw (267.97,269.17) node [anchor=north west][inner sep=0.75pt]  [font=\Large,rotate=-290]  {$l\left( x_{h}^{*}( 1|x_{k}) ,u^*_h( 1|x_{k})\right)$};
\draw (410.97,266.17) node [anchor=north west][inner sep=0.75pt]  [font=\Large,rotate=-290]  {$l\left( x_{h}^{*}\left( N-\tilde{N} |x_{k}\right) ,u^*_h\left( N-\tilde{N} |x_{k}\right)\right)$};
\draw (300.43,611.11) node [anchor=north west][inner sep=0.75pt]  [font=\Large,color={rgb, 255:red, 174; green, 14; blue, 34 }  ,opacity=1 ,rotate=-70]  {$l\left( x_{h}^{*}( 1|x_{k}) ,u_h^*( 1|x_{k})\right)$};
\draw (448.9,609.73) node [anchor=north west][inner sep=0.75pt]  [font=\Large,color={rgb, 255:red, 174; green, 14; blue, 34 }  ,opacity=1 ,rotate=-70]  {$l\left( x_{h}^{*}\left( N-\tilde{N} |x_{k}\right) ,u_h^*\left( N-\tilde{N} |x_{k}\right)\right)$};
\draw (527.33,148.6) node [anchor=north west][inner sep=0.75pt]  [font=\Large]  {$V_{\tilde{N} -1}\left( x_{h}^{*}\left( N-\tilde{N} +1|x_{k}\right)\right)$};
\draw (596.67,709.93) node [anchor=north west][inner sep=0.75pt]  [font=\Large,color={rgb, 255:red, 174; green, 14; blue, 34 }  ,opacity=1 ]  {$V_{\tilde{N}}^{\tilde{N}}\left( x_{h}^{*}\left( N-\tilde{N} +1|x_{k}\right)\right)$};
\draw (471.59,584.06) node [anchor=north west][inner sep=0.75pt]  [font=\Large,color={rgb, 255:red, 174; green, 14; blue, 34 }  ,opacity=1 ,rotate=-290]  {$x^*_{h}\left( N-\tilde{N} +1|x_{k}\right)$};
\draw (353.26,582.39) node [anchor=north west][inner sep=0.75pt]  [font=\Large,color={rgb, 255:red, 174; green, 14; blue, 34 }  ,opacity=1 ,rotate=-290]  {$x^*_{h}\left( N-\tilde{N} |x_{k}\right)$};
\draw (228.62,584.54) node [anchor=north west][inner sep=0.75pt]  [font=\Large,color={rgb, 255:red, 174; green, 14; blue, 34 }  ,opacity=1 ,rotate=-299.93]  {$x^*_{h}( 1|x_{k})$};

\end{tikzpicture}

    }
    \caption{Black: $V_{N}^{\tilde{N}}(x_k)$ is decomposed along the optimal trajectory of Problem \eqref{eq:VN_Ntilde}. It consists of the running costs $l(\cdot,\cdot)$ up to the transition reaching $\textcolor{black}{x_s=x_h^*(N-\tilde{N}+1|x_k)}$, which is the last predicted state required to lie in $\mathcal{X}_1$, followed by the auxiliary tail cost $V_{\tilde{N}-1}(x_s)$. States are shown below the nodes. Blue: The double-headed arrow indicates the equality between the corresponding black and red states. Red: An upper-bound of $V_{N}^{\tilde{N}}(x_{k+1})$ is derived using $x_{k+1}=x^*_h(1|x_k)$, built by the sum of admissible running costs $l(\cdot,\cdot)$ until the state $x_s = x^*_h(N-\tilde{N}+1|x_k)$. Relative to the shifted problem initialized at $x_{k+1}$, this is the penultimate state subject to $\mathcal{X}_1$. As the starting state now is $x^*_h(1|x_k)$ and the constraint horizon associated with $\mathcal{X}_1$ is also $N-\tilde{N}$, the state $x^*_h(1|x_s)$ is the last state required to lie in $\mathcal{X}_1$. The ``tail" of the value function can be computed by $V_{\tilde{N}}^{\tilde{N}}(x_s)$. Dashed black/red arrows have identical costs and cancel in $V^{\tilde{N}}_N(x_k)-V^{\tilde{N}}_N(x_{k+1})$.}
    \label{fig:canIdea}
\end{figure}

\textcolor{black}{Note that, as in other papers estimating suboptimality via RDP (e.g., \cite{grune2008infinite}), non-applicable values $\alpha<0$ may arise.}

\textcolor{black}{
Although informative, \eqref{opt_alpha} does not explicitly show how
$\alpha$ depends on $(N,\tilde N)$. We characterize this dependence next.}

\subsection{Explicit horizon dependence of $\alpha$} \label{BP}
From now on, to ease notation we use 
\begin{equation}
    x_s \coloneqq x^*_{h}(N-\tilde{N}+1|x_k),
\end{equation}
and rewrite $V_{\tilde{N}}^{\tilde{N}}(x^*_{h}(N-\tilde{N}+1|x_k)) - V_{\tilde{N}-1}(x^*_{h}(N-\tilde{N}+1|x_k))$ in \eqref{eq:upbound_dif} as:
\textcolor{black}{
\begin{align} \label{eq:diff2}
& V_{\tilde{N}}^{\tilde{N}}(x_s) - V_{\tilde{N}-1}(x_s) \nonumber \\
& = V_{\tilde{N}}^{\tilde{N}}(x_s) - V_{\tilde{N}}(x_s) + V_{\tilde{N}}(x_s) - V_{\tilde{N}-1}(x_s). 
\end{align}
}
The first difference after the equality is the cost gap between \eqref{eq:VN_Ntilde} and \eqref{eq:VN} at prediction horizon $\tilde{N}$, or the marginal cost of changing the initial constraint. \textcolor{black}{In this work recursive feasibility is not obtained by this difference, nor by choosing a sufficiently long prediction horizon as in \cite{boccia2014stability}, but follows from the imposed invariance of $\mathcal X_1$ and the inclusion $\mathcal X_1\subseteq\mathcal X_2$. Nonetheless, relating invariance to horizon length as in \cite{boccia2014stability} is an interesting future direction.} The second difference is the gap between \eqref{eq:VN} at prediction horizons $\tilde{N}$ and $\tilde{N}-1$, or the marginal cost of adding one prediction step to a problem with initial prediction horizon $\tilde{N}-1$. \textcolor{black}{This difference is in line with suboptimality analyses for MPC without terminal ingredients \cite{worthmann2011stability,grune2010analysis}. A related but distinct mechanism appears in control-horizon MPC: applying $m$ open-loop inputs exposes the residual gap $V_N(x_m)-V_{N-m}(x_m)$, which is bounded via controllability/RDP estimates \cite{worthmann2011stability,grune2010analysis}. In contrast, in \eqref{eq:diff2}, the heterogeneous constraint arrangement makes the first $N-\tilde N$ running costs common to the compared decompositions, allowing these terms to cancel algebraically.} We begin with the first difference in \eqref{eq:diff2}: 
    \begin{subequations}
     \begin{align}
        & V_{\tilde{N}}^{\tilde{N}}(x_s) = \sum^{\tilde{N}-1}_{n=0}l(x^*_{h}(n|x_s),u^*_h(n|x_s)), \\
        & V_{\tilde{N}}(x_s) = \sum^{\tilde{N}-1}_{n=0}l(x^*_{d}(n|x_s),u^*_d(n|x_s)).
    \end{align}
    \end{subequations}
\textcolor{black}{
For the fixed horizon pair $(N,\tilde N)$, we use a single finite constant
$\delta\ge0$, uniform with respect to $x_k\in\mathcal X_1$, such that
\begin{align}\label{eq:VtildeVdiff1}
    V_{\tilde{N}}^{\tilde{N}}(x_s) - V_{\tilde{N}}(x_s)
    \leq
    \delta
    \sum^{\tilde{N}-1}_{n=0}
    l(x^*_{d}(n|x_s),u^*_d(n|x_s)),
    \: \forall x_k\in\mathcal X_1.
\end{align}
}
\noindent 
\textcolor{black}{At \(x_s=0\), both value functions vanish and the above bound holds trivially. Hence, we limit ourselves to starting states \(x_s\neq0\).}
\textcolor{black}{In the considered case, \eqref{eq:VtildeVdiff1} isolates the relative cost of
tightening the initial constraint from $\mathcal X_2$ to $\mathcal X_1$. Thus, $\delta$ separates the constraint-induced value-function gap from the prediction-horizon extension term, enabling \eqref{opt_alpha} to yield an explicit certificate.} \textcolor{black}{To derive it, we impose the following cost-controllability condition, adapted from \cite[Assumption 6.4]{grune2017nonlinear}, which does not imply \emph{exponential state controllability} \cite[Example 6.5]{grune2017nonlinear}.}

\textcolor{black}{
\begin{assumption}[Maximum rate of cost controllability]\label{ass2}
For fixed horizons $N\geq 2$ and $2\leq \tilde N\leq N$ there exist
constants $C_1\geq 1$, $C_2\geq0$, and decay rates
$\sigma_1,\sigma_2\in[0,1)$ such that, for every $x_k\in\mathcal X_1$, the
selected optimal trajectory of \eqref{eq:VN_Ntilde} satisfies
\begin{align}
& l(x^*_{h}(n|x_k),u^*_h(n|x_k))
\leq C_1 \sigma^n_1 \lambda_0,
\label{bctr}\\
& n=0,\dots,N-\tilde{N}. \nonumber
\end{align}
Moreover, the auxiliary tail value functions evaluated below are finite,
attain their minima, and satisfy, for $g\geq 1$,
\begin{align}
& V_g(x^*_{d}(n|x_s))
\leq
C_2 \lambda_{N-\tilde N}
\sum_{i=n+1}^{n+g}\sigma_2^i, \label{bunctr}
\end{align}
\end{assumption}
}
\textcolor{black}{
In Assumption~\ref{ass2} we use an exponential $\mathcal{KL}_0$ function, although more general ones could be used. This assumption may appear restrictive, but it only imposes exponential envelopes on the constrained running costs and on the auxiliary
$\mathcal X_2$-constrained tail value functions. The two
regimes are connected through $\lambda_{N-\tilde N}$, used as a ``bridge quantity": the auxiliary tail is first bounded relative to
$\lambda_{N-\tilde N}$, and this bound is then propagated back to
$\lambda_0$. This does not require monotonic decay of the optimal trajectory, as transient increases can be absorbed by $C_1$ and $C_2$, although large values may yield conservative or vacuous bounds. The tail estimate \eqref{bunctr} is used below with $(n,g)=(0,\tilde N)$
and with $(n,g)=(\tilde N-1,1)$. In the latter case,
$x_d^*(\tilde N-1|x_s)$ denotes the terminal predicted state of an
optimizer of $V_{\tilde N-1}(x_s)$. The following Lemma bounds the second
difference in \eqref{eq:diff2}.
\begin{lemma} \label{lem:sec_bound}
Let Assumption~3 hold. For $N\geq3$, $2\leq\tilde N\leq N-1$, the second
difference in \eqref{eq:diff2} satisfies
\begin{equation} \label{eq:diff_pt2}
    V_{\tilde{N}}(x_s) - V_{\tilde{N}-1}(x_s)
    \leq
    C_1C_2\sigma^{N-\tilde{N}}_1
    \sigma^{\tilde{N}}_2\lambda_0.
\end{equation}
\end{lemma}
Equation \eqref{eq:VtildeVdiff1} and Lemma \ref{lem:sec_bound} bound the two differences in \eqref{eq:diff2}, giving the following result.}
\begin{theorem} \label{thm1} \textcolor{black}{Let Assumption \ref{ass2} hold, and let there exist a single finite constant $\delta\ge0$ satisfying \eqref{eq:VtildeVdiff1} uniformly with respect to $x_k\in\mathcal X_1$}. Then, for  $N \geq 3, 2 \leq \tilde{N} \leq N-1$ and for all $x_k\in\mathcal{X}_1$, \textcolor{black}{an estimate of $\alpha$ in \eqref{eq:alpha}, with explicit dependence on $N$ and $\tilde{N}$, is:} 
    \begin{equation} \label{eq:thm_alpha}
        \alpha =  1- C_1C_2\sigma^{N-\tilde{N}}_1 \sigma_2 \left ( \delta \frac{1-\sigma^{\tilde{N}}_2}{1-\sigma_2} + \sigma_2^{\tilde{N}-1} \right).
    \end{equation}
    \textcolor{black}{If the value of $\alpha$ defined above belongs to $[0,1]$, then it is an admissible RDP certificate.}
\end{theorem}

Unlike \eqref{opt_alpha}, this bound yields a closed-form $\alpha$ as a function of $N$, $\tilde{N}$, $C_1$, $C_2$, $\sigma_1$, $\sigma_2$, $\delta$. If $V_{\tilde{N}}(x_s)$ automatically produces $x^*_{d}(1|x_s) \in \mathcal{X}_1$, there is no need to calculate $V_{\tilde{N}}^{\tilde{N}}(x_s)$, as $V_{\tilde{N}}^{\tilde{N}}(x_s) = V_{\tilde{N}}(x_s)$, implying $\delta$ is not needed, and $\alpha$ in Theorem \ref{thm1} reduces to:  
\begin{equation} \label{eq:alp_red}
\alpha = 1- C_1C_2\sigma^{N-\tilde{N}}_1 \sigma_2^{\tilde{N}}.
\end{equation}

The next Proposition shows how Assumption \ref{ass2} provides an upper bound for $J^{N,\tilde{N}}_{\infty}(x_k)$ \eqref{eq:alpha} based on decay rates and $\alpha$.   
\begin{proposition} \label{corol1}
 \textcolor{black}{Let Assumption \ref{ass2} hold} and $\alpha \in (0,1]$. Then by \eqref{bctr}, \eqref{bunctr} an upper bound for $J^{N,\tilde{N}}_{\infty}(x_k)$ is:
\begin{align} \label{eq:cl_ub}
   & J^{N,\tilde{N}}_{\infty}(x_k) \leq C_1 \left [\left(\frac{1-\sigma_1^{N-\tilde{N}+1}}{1-\sigma_1} \right) + \right. \nonumber \\ 
   & \left. C_2\sigma_1^{N-\tilde{N}}\sigma_2 \left(\frac{1-\sigma_2^{\tilde{N}-1}}{1-\sigma_2} \right) \right] \frac{\max_{u \in \mathcal{U}} l(x_k,u)}{\alpha}.
\end{align}
\end{proposition}
\textcolor{black}{The parameter $\alpha$ used in \eqref{eq:cl_ub} can be obtained  from either \eqref{opt_alpha} or  Theorem~\ref{thm1}. Theorem~\ref{thm1} provides an upper bound expressed only in terms of estimated parameters, clarifying the effect of $N$ and $\tilde{N}$ on the closed-loop upper-bound.}
\textcolor{black}{Estimating \(\delta\) directly from \eqref{eq:VtildeVdiff1} is demanding, since it requires the additional evaluation of \(V_{\tilde N}(x_s)\), beyond
the quantities \(V^{\tilde N}_{\tilde N}(x_s)\) and
\(V^{\tilde N}_{N}(x_k)\) already involved in the certificate.
We next discuss explicit ways of computing or upper-bounding
\(\delta\).}

\subsubsection{Discussion on $\delta$} \textcolor{black}{ Certified bounds using Theorem \ref{thm1} require either direct evaluation of $\delta$ by \eqref{eq:VtildeVdiff1} or via a more conservative, but less computationally intensive, construction.}
\textcolor{black}{
\begin{proposition}\label{prop:delta_bd}
For each \(x_k\in\mathcal X_1\setminus\{0\}\), let
\(x_s\neq0\), and let
\(x_d^*(\cdot|x_s),u_d^*(\cdot|x_s)\) be an optimizer of
\(V_{\tilde N}(x_s)\). Suppose that, for every such \(x_k\), there exist
\(z_s\in\mathcal X_1\) and \(u_0,u_1\in\mathcal U\) such that
\(z_s=f(x_s,u_0)\) and \(x_d^*(2|x_s)=f(z_s,u_1)\). Suppose further that
there exist constants \(\varepsilon>0\) and \(c\geq1\), independent of
\(x_k\), such that, whenever \(0<\|x_s\|\leq\varepsilon\),
\begin{align}\label{eq:eps}
l(x_s,u_0)+l(z_s,u_1)
\leq
c\min_{v\in\mathcal U}l(x_d^*(0|x_s),v).
\end{align}
Then \eqref{eq:VtildeVdiff1} holds uniformly over \(\mathcal X_1\) with
\begin{align}\label{eq:deltaeq}
\delta
:=
\max\left\{
c-1,\,
\sup_{\substack{x_k\in\mathcal X_1\setminus\{0\}\\
\|x_s\|\geq\varepsilon}}
\left[
\frac{l(x_s,u_0)+l(z_s,u_1)}
{\min_{v\in\mathcal U}l(x_d^*(0|x_s),v)}
-1
\right]
\right\}.
\end{align}
The split in \eqref{eq:deltaeq} improves numerical conditioning. As
$x_s\to0$, both the numerator and denominator vanish, so \eqref{eq:eps}
can be verified locally to prevent the ratio from becoming unbounded.
For $\|x_s\|\geq\varepsilon$, the denominator is bounded away from zero,
and the remaining supremum is finite by compactness, continuity, and
positive definiteness. If a closed-form constant $c$ satisfying
\eqref{eq:eps} is available for all admissible $x_s\neq0$, set
$\delta=c-1$ directly. In particular, if $z_s=x_s$,
$l(x,u)=x^\top Qx+u^\top Ru$, and $0\in\mathcal U$, then the same
conclusion holds with
\begin{align*}
\delta
:=
\max\left\{
\underbrace{c-1}_{0<\|x_s\|\leq\varepsilon},\,
\underbrace{
1+\frac{2\max_{u\in\mathcal U}u^\top Ru}
{\lambda_{\min}(Q)\varepsilon^2}
}_{\|x_s\|\geq\varepsilon}
\right\}.
\end{align*}
\end{proposition}
}
\textcolor{black}{The certificate in \eqref{eq:deltaeq} provides a general way of
upper-bounding the value function mismatch in
\eqref{eq:VtildeVdiff1}. Its verification may require solving a nonlinear
reachability problem and be comparable in difficulty to evaluating
\eqref{eq:VtildeVdiff1} directly. The specialization $z_s=x_s$ with a
quadratic cost is more restrictive but, when applicable, simplifies
verification of the local constant $c$ and yields a conservative explicit
expression for $\delta$. Once verified, \eqref{eq:deltaeq} avoids the
direct computation of $V^{\tilde N}_{\tilde N}(x_s)$.}

\subsubsection{\textcolor{black}{Stability considerations via horizon selection}}
\textcolor{black}{As previously explained, $\alpha>0$ certifies asymptotic stability, and $\alpha = 0$ gives a degenerate suboptimality bound, but certifies stability}. We now study how $N,\tilde{N}$ affect $\alpha$ via Theorem \ref{thm1} \textcolor{black}{to conclude closed-loop
stability}. \textcolor{black}{For Theorem \ref{thm1}, Assumption \ref{ass2} only needs to hold uniformly over $x_k\in\mathcal X_1$ for a fixed pair $(N,\tilde N)$. Here, since the next result is used as an \emph{a priori} horizon-selection condition for closed-loop
stability, we additionally require the bounds in Assumption~\ref{ass2} to hold
uniformly over all admissible horizon pairs.}
\begin{proposition}\label{prop_Nstab}
Assume that $C_1,C_2>0,\sigma_1,\sigma_2$, and $\delta$ are known and valid
uniformly over all admissible horizon pairs $(N,\tilde N)$ and all $x_k\in\mathcal X_1$. \textcolor{black}{Assume also that $V_N^{\tilde N}(\cdot)$ is
continuous at the origin for every admissible horizon pair.} Let
$\sigma=\max\{\sigma_1,\sigma_2\}$. If $\sigma=0$, then $\alpha=1$.
Otherwise, if $\sigma\in(0,1)$, any horizons satisfying $N\geq3$,
$2\leq\tilde N\leq N-1$, and
\begin{equation}\label{eq:stabhoriz}
    N-\tilde N+1
    \geq
    \left\lceil
    \frac{
    \log\left(C_1C_2\left[\frac{\delta}{1-\sigma_2}+1\right]\right)
    }{
    \log\left(\frac{1}{\sigma}\right)
    }
    \right\rceil
\end{equation}
guarantee $\alpha\geq0$ in Theorem \ref{thm1}, and hence certify stability of
the corresponding closed-loop.
\end{proposition}
\textcolor{black}{Uniformly verifying Assumption~\ref{ass2} remains an open point. Arguments similar to \cite[Proposition~4.7]{grune2008infinite} may provide a useful starting point, although the heterogeneous constraints
and auxiliary value functions in our setting require additional care.} \textcolor{black}{ Proposition \ref{prop_Nstab} shows that stability (and a suboptimality bound) can be certified for suitable choices of $N$ and $\tilde N$, based on the system’s decay-rate characteristics. Relation \eqref{eq:stabhoriz} is a simplified, closed form, sufficient condition for stability. A more complex implicit expression depends on both horizons separately.}

\section{Closed-loop performance lower bound} \label{sec:cl_lbound}
\subsection{Calculation of $\omega$ as a difference of \textcolor{black}{value} functions}
We revisit \eqref{eq:clbd} in Proposition \ref{prop_lb}, as we aim to estimate $\omega$ as a function of $(N,\tilde{N})$. Now, we do so by deriving an upper bound for $V_N^{\tilde N}(x_k)- V_N^{\tilde N}(x_{k+1})$, and constructing an inequality chain using \eqref{eq:dyn_prog_2}. We again transform $V_N^{\tilde N}(x_k)- V_N^{\tilde N}(x_{k+1})$ into a difference of value functions; the following Lemmas are useful.  
\begin{lemma}\label{lem:eqVNNtilde}
     \textcolor{black}{Let Assumption~\ref{wellpd} hold.} Then
    \textcolor{black}{
    \begin{align}
    & V^{\tilde{N}}_{\tilde{N}}(x^*_{h}(N-\tilde{N}|x_{k+1})) \nonumber \\ 
    & = l(x^*_h(N-\tilde{N}|x_{k+1}),u^*_h(N-\tilde{N}|x_{k+1})) \nonumber \\
    & + \sum^{N-1}_{n=N-\tilde{N}+1}l(x^*_h(n|x_{k+1}),u^*_h(n|x_{k+1})).
    \end{align}
    }
\end{lemma}
\textcolor{black}{
The next lemma provides the upper-bound counterpart to the lower bound in Lemma~4. Both lemmas bound the same decrement \(V_N^{\tilde N}(x_k)-V_N^{\tilde N}(x_{k+1})\), but from opposite sides: Lemma~4 gives the lower bound used for the upper-RDP certificate, whereas Lemma \ref{lem:diff_val_func_lbd} below gives the upper bound used for the lower-RDP certificate. Their structural similarity follows from the same shifted-trajectory argument, applied at the corresponding switching states.
}
\begin{lemma}\label{lem:diff_val_func_lbd} \textcolor{black}{Let Assumption~\ref{wellpd} hold.} For $N \geq 3, 2 \leq \tilde{N} \leq N-1$, we can upper-bound $V_N^{\tilde N}(x_k) - V_N^{\tilde N}(\textcolor{black}{x_{k+1}})$ as:
    \begin{align}\label{eq:diff_val_func_lbd}
        & V_N^{\tilde N}(x_k) - V_N^{\tilde N}(\textcolor{black}{x_{k+1}})\le \textcolor{black}{\lambda_0} \nonumber  \\
        & -\left (V_{\tilde{N}}^{\tilde{N}}(x^*_{h}(N-\tilde{N}|x_{k+1})) - V_{\tilde{N}-1}(x^*_{h}(N-\tilde{N}|x_{k+1})) \right).
    \end{align}
\end{lemma}
\textcolor{black}{Consistent with the discussion following Lemma \ref{lem:diff_val_func_ubd},
subsequent results using \eqref{eq:diff_val_func_lbd} retain the horizon restrictions of Lemma \ref{lem:diff_val_func_lbd}.
For $\tilde N=N$, $x_h^*(0\mid x_{k+1})=x_{k+1}$, \eqref{eq:diff_val_func_lbd} becomes an equality, and obtaining an admissible $\omega$ reduces to verifying \eqref{eq:dyn_prog_2}.
This boundary case is also not pursued.}
Using Lemmas \ref{lem:eqVNNtilde}, \ref{lem:diff_val_func_lbd}, and having \eqref{eq:dyn_prog_2} as a goal, we enforce the inequality chain below (valid for $\alpha \textcolor{black}{>}0$): 
    \begin{align} \label{eq:upbound_diff_k+1}
        & \frac{\alpha}{1-\omega} [V_N^{\tilde N}(x_k) - V_N^{\tilde N}(\textcolor{black}{x_{k+1}})]\le \frac{\alpha}{1-\omega} \left[\textcolor{black}{\lambda_0} \nonumber \right. \\
        & \left. -\left (V_{\tilde{N}}^{\tilde{N}}(x^*_{h}(N-\tilde{N}|x_{k+1})) - V_{\tilde{N}-1}(x^*_{h}(N-\tilde{N}|x_{k+1})) \right) \right]\nonumber \\
        & \leq \alpha \textcolor{black}{\lambda_0},
    \end{align}

\textcolor{black}{
We can now state our implicit 
$\omega$ estimate.
\begin{lemma}\label{lem:implicit_omega}
For \(N\ge 3\) and \(2\le \tilde N\le N-1\), suppose that, for all 
\(x_k\in\mathcal X_1\) with \(x_{k+1}\) obtained via 
\eqref{eq:closed-loop-dynamics}, the intermediate upper bound in 
\eqref{eq:upbound_diff_k+1} is nonnegative. Then, a uniform lower-RDP certificate can be obtained by
\begin{align} \label{opt_omega}
    \omega & =  \inf_{x_k\in\mathcal X_1\setminus\{0\}}\frac{1}{\lambda_0} \left[ V_{\tilde{N}}^{\tilde{N}}(x^*_{h}(N-\tilde{N}|x_{k+1})) \right. \nonumber \\
    & \left. - V_{\tilde{N}-1}(x^*_{h}(N-\tilde{N}|x_{k+1})) \right].
\end{align}
If $\omega\in[0,1)$, then \eqref{eq:dyn_prog_2} holds for all 
$x_k\in\mathcal X_1$.
\end{lemma}
}
\textcolor{black}{Compared to the direct verification of
\eqref{eq:dyn_prog_2}, the implicit certificate in \eqref{opt_omega}
rewrites the value decrement as a tail value function difference involving horizons
\(\tilde N\) and \(\tilde N-1\). Unlike the implicit upper-bound certificate
in Lemma~\ref{lem:implicit_alpha}, however, this does not generally remove
the need to characterize the optimizer initialized at \(x_{k+1}\). Thus,
the main role of \eqref{opt_omega} is analytical: it motivates the
horizon-explicit certificate derived next. We highlight that even if $\omega=0$, it certifies that
under the presented} circumstances, $V^{\tilde{N}}_{N}(x_k)$ is a
reasonable lower-bound choice.

\textcolor{black}{Note that $\omega$ is not computed
from the numerical value of $\alpha$. When $\alpha>0$, the lower-RDP
condition \eqref{eq:dyn_prog_2} is equivalent to
$V_N^{\tilde N}(x_k)-V_N^{\tilde N}(x_{k+1})
\leq(1-\omega)\lambda_0$. Thus, admissible choices of $\omega$ are obtained
from an upper bound on this value-function difference. Nevertheless, the
resulting closed-loop lower bound is nontrivial only when a positive
upper-RDP certificate $\alpha>0$ is also available. If $\alpha=0$,
Proposition~\ref{prop_lb} reduces to the trivial inequality.}

\textcolor{black}{As with the other implicit bounds, the calculation of some
value functions is required. $V_{N}^{\tilde{N}}(x_k)$ is needed to obtain
$\lambda_0$,} \textcolor{black}{while}
\textcolor{black}{$V_{\tilde{N}-1}
(x^*_{h}(N-\tilde{N}|x_{k+1}))$}
\textcolor{black}{must be computed separately.}
\textcolor{black}{This certificate does not make explicit how this estimate
varies} with the parameters $N$, $\tilde{N}$. As before, we proceed to
characterize $\omega$ explicitly in terms of these parameters.

\subsection{Explicit horizon dependence of $\omega$}

\textcolor{black}{We henceforth denote 
\begin{equation}
    x_p \coloneqq x^*_{h}(N-\tilde{N}|x_{k+1})
\end{equation}
to lighten notation.} Mirroring \eqref{eq:diff2}, we re-write the bound with $x_p$ rather than $x_h^*(N-\tilde{N}|x_{k+1})$:
\begin{align} \label{eq:diff3}
& V^{\tilde N}_{\tilde N}(x_p) - V_{\tilde{N}-1}(x_p) \nonumber \\
& = V_{\tilde{N}}^{\tilde{N}}(x_p) - V_{\tilde{N}}(x_p) + V_{\tilde{N}}(x_p) - V_{\tilde{N}-1}(x_p) \nonumber \\
& \geq  V_{\tilde{N}}(x_p) - V_{\tilde{N}-1}(x_p).
\end{align} 

The same considerations about marginal costs made for \eqref{eq:diff2} are also valid here. \textcolor{black}{In contrast to \eqref{eq:VtildeVdiff1}, we discard the first value-function difference. Analogous results to those in \eqref{eq:VtildeVdiff1} could also be derived here, leading to the following lower-bounding relation: For a fixed horizon pair $(N,\tilde N)$, we could use a single finite constant $\nu \geq 0$, uniform with respect to $x_k\in\mathcal X_1$, such that:
\begin{align}\label{eq:VtildeVdiff_lowernu_bound_mod}
    V_{\tilde{N}}^{\tilde{N}}(x_p) - V_{\tilde{N}}(x_p)
    \geq
    \nu \sum^{\tilde{N}-1}_{n=0}
    l(x^*_{d}(n|x_p),u^*_d(n|x_p)),
    \: \forall x_k\in\mathcal X_1 .
\end{align}
}
\textcolor{black}{
There exists nonetheless a distinction from \eqref{eq:VtildeVdiff1}. If the constraint present in $V_{\tilde{N}}^{\tilde{N}}(x_s)$ but absent from $V_{\tilde{N}}(x_s)$ is active at least once over $\mathcal{X}_1$, then a nontrivial $\delta>0$ can be obtained. Here, however, the opposite situation is relevant: if the constraint present in $V_{\tilde{N}}^{\tilde{N}}(x_p)$ but absent from $V_{\tilde{N}}(x_p)$ is inactive at least once over $\mathcal{X}_1$, with the corresponding $x_p \neq 0$, then $V_{\tilde{N}}^{\tilde{N}}(x_p)=V_{\tilde{N}}(x_p)$, yielding $\nu=0$. Since this is the typical case, we focus only on $V_{\tilde{N}}(x_p)-V_{\tilde{N}-1}(x_p)$. In the less common cases where a nontrivial $\nu>0$ can be certified, one can instead use \eqref{opt_omega}. We now state an assumption analogous to Assumption \ref{ass2} for our explicit bound.}
\begin{assumption}[\textcolor{black}{Minimum rate of cost controllability}]
\label{ass4}
\textcolor{black}{Consider fixed horizons $N\geq 2$ and
$2\leq\tilde N\leq N$. There exist constants $1\geq C_3>0$,
$C_4>0$, decay rates $\sigma_3,\sigma_4\in(0,1)$ and $\kappa \geq 0$ such that, for every
$x_{k}\in\mathcal X_1$, the selected optimal trajectory of
\eqref{eq:VN_Ntilde} initialized at the corresponding \(x_{k+1}\) satisfies}
\begin{align}
& l(x^*_{h}(n|x_{k+1}),u^*_h(n|x_{k+1}))
\geq C_3 \sigma^n_3 \lambda^+_0,
\label{lctr}\\
&\lambda^+_0
= l(x^*_{h}(0|x_{k+1}),u^*_h(0|x_{k+1})), \nonumber \\
& n=0,\dots,N-\tilde{N}. \nonumber
\end{align}
\textcolor{black}{Moreover, the selected optimal trajectory of
$V_{\tilde N}(x_p)$ satisfies}
\begin{align}
& l(x^*_{d}(n|x_p),u^*_d(n|x_p))
\geq C_4 \sigma^{n+1}_4 \lambda^{+}_{N-\tilde{N}},
\label{blctr}\\
&\lambda^{+}_{N-\tilde{N}}
= l(x^*_{h}(N-\tilde{N}|x_{k+1}),
u^*_h(N-\tilde{N}|x_{k+1})),
\nonumber \\
& n=0,\dots,\tilde{N}-1. \nonumber
\end{align}
\textcolor{black}{Finally, the first closed-loop stage cost at time $k+1$ is assumed to satisfy
\begin{equation} \label{eq:kappa}
    \lambda_0^+\geq \kappa \lambda_0,
\end{equation}}
\end{assumption}

\textcolor{black}{Assumption~\ref{ass4} is a lower-envelope condition on the
selected optimal costs used in the lower-bound construction. Inequality \eqref{eq:kappa} compares the lower-envelope estimate initialized at $x_{k+1}$ with the stage cost $\lambda_0$ incurred at the previous closed-loop step. Although a minimum-rate condition may appear counterintuitive, it is relevant when admissible inputs saturate or when constraints prevent the optimizer from reducing the stage cost
arbitrarily fast. In such cases, the optimal cost decrease may admit a
nontrivial lower exponential envelope.}

\textcolor{black}{We emphasize that Assumption~\ref{ass4} is a sufficient condition for obtaining a lower-bound certificate explicit in the horizons. If the constants required by Assumption~\ref{ass4} cannot be verified, or if the assumption is not satisfied, the value-function difference \eqref{opt_omega} can still be evaluated directly. We now derive a relation for $\omega$ explicit on the horizons $(N,\tilde N)$.}

\begin{theorem} \label{thm2}
\textcolor{black}{Let Assumption \ref{ass4} hold. Then, for
$N \geq 3$, $2 \leq \tilde{N} \leq N-1$, and for all
$x_k\in\mathcal{X}_1$, an admissible estimate of $\omega$ in \eqref{eq:dyn_prog_2}, with explicit dependence on $N$ and $\tilde N$, is}
\begin{equation} \label{eq:thm2}
\omega = C_3 C_4 \kappa \sigma^{N-\tilde{N}}_3 \sigma_4^{\tilde{N}},
\end{equation}
\textcolor{black}{provided that $\omega \in [0,1)$.}
\end{theorem}

\textcolor{black}{The constants
$C_3,C_4,\sigma_3,\sigma_4$ and $\kappa$ satisfying the preceding
bounds are not necessarily unique. Each admissible choice yields a valid
certificate, nonetheless conservative choices of these parameters may lead to a weak or trivial bound. Hence, over the admissible constants satisfying the stated inequalities and $\omega<1$ it is desirable to obtain the parameters producing the tightest lower-bound certificate. We further highlight that in case a one-step convergence is attained at any point in $\mathcal{X}_1$, i.e., $x_{k+1}=0$ for some $x_k\neq 0$, then $\lambda_0^+=0$ while $\lambda_0>0$ by Assumption~1. Hence the only admissible choice is $\kappa=0$, which yields the trivial value $\omega=0$ in Theorem~\ref{thm2}. This is not a limitation of the construction: in this case the value-function evaluation already provides the exact lower bound, so no nontrivial ratio-based improvement through $\omega>0$ is needed.}

\textcolor{black}{
We again emphasize that, when using the same fixed horizon pair $(N,\tilde N)$ on Theorems \ref{thm1} and \ref{thm2}, no compatibility condition between  $\alpha$ and $\omega$ is needed. If both the upper and
lower bound certificates are valid on the same set of initial states then, for \(x_k\neq 0\), they bound the same normalized value-function difference from below and above:}
\[
\textcolor{black}{
\alpha \leq \frac{V_N^{\tilde N}(x_k)- V_N^{\tilde N}(x_{k+1})}{\lambda_0}\leq 1-\omega.}
\]
\textcolor{black}{At \(x_k=0\), zero input yields \(\lambda_0=0\) and \(V_N^{\tilde N}(x_k)-V_N^{\tilde N}(x_{k+1})=0\). Hence the inequalities corresponding to \eqref{eq:dyn_prog} and \eqref{eq:dyn_prog_2} hold trivially at the origin. For \(x_k\neq0\), equivalently \(\lambda_0>0\), the normalized inequality above yields \(\alpha\leq 1-\omega\). If this relation fails for the same horizon pair and state, then at least one of the horizon-explicit certificates is not valid with the chosen constants.} \textcolor{black}{Unlike \eqref{opt_omega}, \eqref{eq:thm2} avoids the direct
computation of \(V_{\tilde N-1}(x_p)\), but still requires auxiliary information from \(V_{\tilde N}(x_p)\) to estimate the constants in Assumption~\ref{ass4}. Therefore, Theorem~\ref{thm2} should be viewed primarily as a horizon-explicit analytical certificate for a fixed pair \((N,\tilde N)\), rather than as a major computational shortcut for
verifying the lower-RDP condition.}

Using Theorem \ref{thm2} with Assumption \ref{ass4}, we obtain in Proposition \ref{corol2}, a lower bound for $J^{N,\tilde{N}}_{\infty}(x_k)$ depending solely on the estimated parameters and $\lambda_0$.
\begin{proposition} \label{corol2}
 \textcolor{black}{Let Assumption \ref{ass4} hold, assume $V^{\tilde N}_{N} (\cdot)$ to be continuous at the origin, and let $\alpha \in(0,1]$ and \(\omega\in[0,1)\) be admissible upper and lower-RDP coefficients respectively.} Then, \textcolor{black}{using \eqref{eq:partial_lb}, \eqref{lctr}, and \eqref{blctr}}, the lower-bound for $J^{N,\tilde{N}}_{\infty}(x_k)$ can be written as:
\begin{align} \label{eq:cl_lb}
   & J^{N,\tilde{N}}_{\infty}(x_k) \geq  C_3 \kappa \left [ \left (\frac{1-\sigma_3^{\textcolor{black}{N-\tilde{N}}}}{1-\sigma_3} \right) + \right. \nonumber \\ 
   & \left. C_4\sigma_3^{N-\tilde{N}}\sigma_4 \left(\frac{1-\sigma_4^{\textcolor{black}{\tilde{N}}}}{1-\sigma_4} \right) \right] \frac{\min_{u \in \mathcal{U}} l(x_k,u)}{1-\omega}.
\end{align}
\end{proposition}
From \eqref{eq:cl_lb} and the positive definiteness of $l(\cdot,\cdot)$, $\min_{u \in \mathcal{U}} l(x_k,u) = 0$ only if $x_k=0$. Although any estimated $\omega$ can be used in Proposition \ref{corol2}, using Theorem \ref{thm2} provides a lower-bound based on estimated parameters and $\lambda_0$ alone. 

\section{Numerical simulations}
\label{sec:simulations}

\textcolor{black}{The simulations below illustrate two uses of the proposed bounds. The first is an \emph{a priori} certification, where the parameters are verified uniformly over $\mathcal X_1$. The second is an \emph{a posteriori} trajectory-based evaluation, where the parameters are extracted along the simulated closed-loop trajectory, yielding a tighter and computationally cheaper certificate. Nonetheless, this certificate depends on the considered initial condition and is not uniform over $\mathcal X_1$.}

\subsection{A priori example}

\textcolor{black}{
We start by showing an example where parameters used in Theorems \ref{thm1} and \ref{thm2} are uniformly calculated over the control invariant set $\mathcal{X}_1$. We consider the unstable scalar system
\begin{equation}
    x_{k+1}=\frac{3}{2}x_k+u_k,
    \qquad
    l(x_k,u_k)=x_k^2+16u_k^2,
\end{equation}
with
\[
    \mathcal X_1=[-5,5],\qquad
    \mathcal X_2=[-15,15],\qquad
    \mathcal U=[-5,5].
\]
Assumption~2 holds since \(u=-x/2\) is admissible for every
\(x\in\mathcal X_1\) and gives \(f(x,u)=x\in\mathcal X_1\).
Moreover, \(\mathcal X_1\subset\mathcal X_2\), and \(l\) is continuous
and positive definite. For fixed \(N\) and \(\tilde{N}\),
Problem~(2) has an explicit piecewise affine solution in \(x_k\) \cite{bemporad2000explicit}, while the stage costs and value functions are piecewise quadratic.
}
\subsubsection{Implicit certificates}
\begin{figure}[t]
    \centering
    \begin{minipage}{0.43\textwidth}
        \centering
        \includegraphics[width=\linewidth]
        {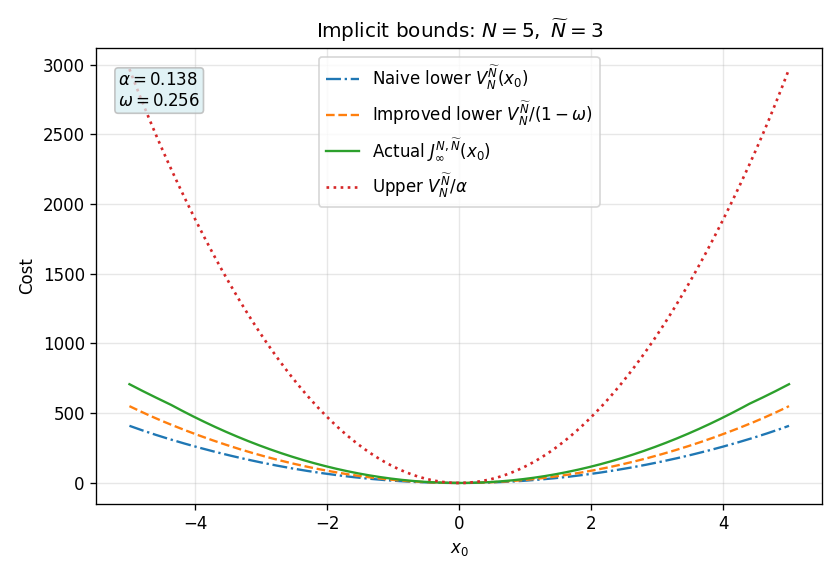}

        \vspace{0.4ex}
        \small
        \textbf{(a)} \(N=5\), \(\widetilde{N}=3\):
        \(\alpha = 0.138\) and
        \(\omega = 0.256\).
    \end{minipage}

    \vspace{1.2ex}

    \begin{minipage}{0.43\textwidth}
        \centering
        \includegraphics[width=\linewidth]
        {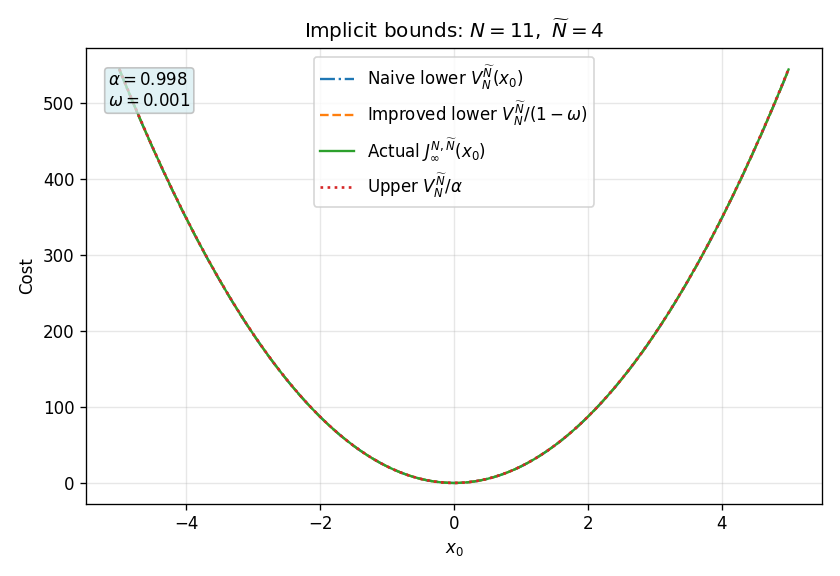}

        \vspace{0.4ex}
        \small
        \textbf{(b)} \(N=11\), \(\widetilde{N}=4\):
        \(\alpha = 0.998\) and
        \(\omega = 0.001\).
    \end{minipage}

    \caption{Uniform implicit bounds obtained directly from the
    value-function differences in (17) and (34). The positive values of
    \(\omega\) provide a noticeable and a marginal improvement over
    \(V_N^{\widetilde{N}}(\cdot)\), for (a) and (b) respectively.}
    \label{fig:implicit-scalar-comparison}
\end{figure}

\textcolor{black}{
Following \eqref{opt_alpha} and \eqref{opt_omega}, respectively, the value-function differences are taken over the full set \(\mathcal X_1\).
For \((N,\widetilde{N})=(5,3)\), the uniform calculation gives $\alpha=0.138, \: \omega=0.256$, while for \((N,\widetilde{N})=(11,4)\), we obtain $\alpha=0.998, \: \omega=0.001$. Both calculated values of $\omega$ are strictly positive, therefore improving the naive lower bound $V^{\tilde{N}}_N(\cdot)$. In particular, for \((N,\widetilde{N})=(5,3)\), the improved lower bound is approximately $34.4\%$ higher, compared with the naive bound \(V_5^3(x_k)\). The closed-loop bound evaluated according to \eqref{eq:clbd}, the actual closed-loop cost and naive bound for these examples are shown in Figure \ref{fig:implicit-scalar-comparison}.
}

\subsubsection{Horizon-explicit certificates}

\textcolor{black}{
For each fixed pair $(N,\widetilde{N})$ and fixed decay rates
$\sigma_i\in(0,1)$, the constants are computed from the active-set
partition of the corresponding parametric quadratic programs. On each
active-set interval, the optimizer is affine in $x_k$, so the ratios
below are rational functions. Their extrema are obtained by evaluating
the corresponding interval endpoints and stationary points. This yields
the smallest $C_1$ and $C_2$ constants satisfying the upper-decay inequalities in Assumption~3, and the largest $C_3$ and $C_4$ constants satisfying the lower-decay inequalities in
Assumption~4. These constants are uniformly calculated over $\mathcal X_1$ as:
\begin{align}
    C_1
    &=
    \max_{\substack{x_k\in\mathcal X_1\setminus\{0\}\\
                    0\leq n\leq N-\widetilde{N}}}
    \frac{
        l\left(x_h^*(n|x_k),u_h^*(n|x_k)\right)}
    {
        \sigma_1^n
        l\left(x_k,u_h^*(0|x_k)\right)},\\
    C_2
    &=
    \max_{\substack{x_k\in\mathcal X_1\setminus\{0\}\\
                    0\leq n\leq \widetilde{N}-1}}
    \frac{1}{\left(\sum_{j=n}^{\widetilde{N}-1}\sigma_2^{j+1}\right)} \nonumber \\
    & \times \frac{V_{\widetilde{N}-n}(x_d^*(n|x_s))}
    {
    l\left(
        x_h^*(N-\widetilde{N}|x_k),
        u_h^*(N-\widetilde{N}|x_k)
    \right)}, \\
    C_3
    &=
    \min_{\substack{x_k\in\mathcal X_1\setminus\{0\}\\
                    0\leq n\leq N-\widetilde{N}}}
    \frac{
        l\left(x_h^*(n|x_{k+1}),u_h^*(n|x_{k+1})\right)}
    {
        \sigma_3^n
        l\left(x_{k+1},u_h^*(0|x_{k+1})\right)},\\
    C_4
    &=
    \min_{\substack{x_k\in\mathcal X_1\setminus\{0\}\\
                    0\leq n\leq \widetilde{N}-1}}
    \frac{
        l\left(x_d^*(n|x_p),u_d^*(n|x_p)\right)}
    {
        \sigma_4^{n+1}
        l\left(
            x_p,
            u_h^*(N-\widetilde{N}|x_{k+1})
        \right)} .
\end{align}
For this first calculation of $\alpha$ via Theorem 1, $\delta$ is evaluated according to its definition in \eqref{eq:VtildeVdiff1}:
\[
\delta
=
\sup_{x_k\in\mathcal X_1\setminus\{0\}}
\frac{
V_{\widetilde{N}}^{\widetilde{N}}(x_s)
-V_{\widetilde{N}}(x_s)}
{V_{\widetilde{N}}(x_s)}.
\]}
\textcolor{black}{
The constant entering Theorem~2 is computed independently as
\[
\kappa
=
\inf_{x_k\in\mathcal X_1\setminus\{0\}}
\frac{\lambda_0^+}{\lambda_0} .
\]
Hence, \(\delta\) uniformly bounds the normalized mismatch between the
constrained and unconstrained auxiliary value functions, whereas
$\kappa$ is the largest uniform lower bound relating the first stage
costs at $x_{k+1}$ and $x_k$. For $(N,\widetilde{N})=(6,3)$, direct evaluation of \eqref{eq:VtildeVdiff1} gives $\delta=0.014,\: \kappa=0.55$ and the following constants $(C_1,C_2,C_3,C_4) =(1.2,3.7,0.32,0.92)$ and decay rates $(\sigma_1,\sigma_2,\sigma_3,\sigma_4) =(0.42,0.99,0.65,0.99)$, respectively. This produces $\alpha = 0.671444$, and $\omega = 0.043186$. Thus, Theorem~2 increases the naive lower-bound coefficient by approximately $4.51\%$, for a uniform positive $\alpha$ certificate obtained from Theorem 1.}
\begin{figure}[t]
\centering
\includegraphics[width=0.43\textwidth]
{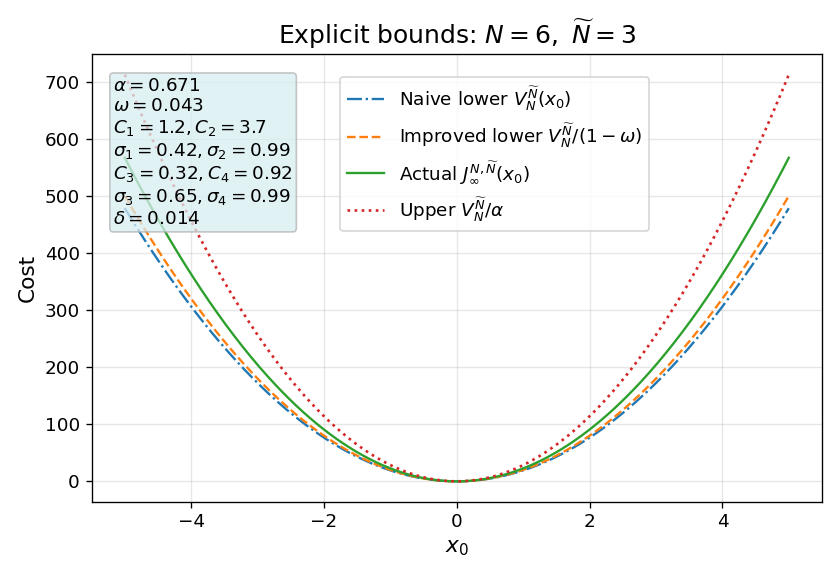}
\caption{Uniform horizon-explicit bounds for
$(N,\widetilde N)=(6,3)$, where $\delta$ is evaluated directly
from its definition in \eqref{eq:VtildeVdiff1}.}
\label{fig:explicit-direct-delta-scalar}
\end{figure}
\textcolor{black}{
For the second explicit calculation of \(\alpha\), \(\delta\) is instead obtained from the one-step holdability specialization in
Proposition \ref{prop:delta_bd}. Since \(u_0=-x_s/2\) gives
\(f(x_s,u_0)=x_s\) uniformly on \(\mathcal X_1\), we can
choose $z_s=x_s$. For a fixed \(\widetilde N=4\) and \(\varepsilon=5\), upper-bounding the ratio in \eqref{eq:eps} over all $x_s\in\mathcal X_1\setminus\{0\}$ gives the $N$-independent bound $c=6.53$, yielding \(\delta=c-1=5.53\). Calculation via the specialization \eqref{eq:deltaeq} instead gives the conservative estimate}
\textcolor{black}{
\[
\delta
=
\max\left\{c-1,\,
1+\frac{2\max_{u\in\mathcal U}16u^2}{\varepsilon^2}\right\}
=33.
\]
We varied \(N\in\{6,\ldots,11\}\). Using \(\delta=5.53\), the first horizon pair yielding
\(\alpha>0\) is \((N,\widetilde N)=(9,4)\). The constants calculated were $(C_1,C_2,C_3,C_4)=(1.01,1.91,0.15,0.47)$, $(\sigma_1,\sigma_2,\sigma_3,\sigma_4)=(0.41,0.99,0.60,0.99)$, \(\kappa=0.41\), producing \(\alpha=0.51\), and \(\omega=0.002148\).
At \((11,4)\) the constants values are approximately equal with the exception of $C_3=0.04$ and $\sigma_3 = 0.64$. The uniform bound gives \(\alpha=0.92\), while if we were to use \(\delta=33\), we would get
\(\alpha=0.54\). As \(\omega\) is independent of \(\delta\), both calculations
give \(\omega=0.000344\).
This example highlights two aspects. First, the implicit value-function difference yields tight $\omega$ estimates for relatively low prediction/constraint horizons, thus noticeably improving the naive lower bound in this region. By contrast, because a nontrivial lower-bound certificate requires $\alpha>0$, the more conservative
combination of Theorem~\ref{thm1} and Proposition~\ref{prop:delta_bd}
requires longer horizons to obtain a positive or comparable $\alpha$, a region
where $\omega$ is typically less informative. Second, uniform bounds over $\mathcal{X}_1$ provide useful but intrinsically conservative a priori design information. Evaluating the bounds along the closed-loop trajectory instead provides a less conservative a posteriori assessment of the controller with respect to the infinite-horizon fully constrained optimal
controller, as shown next.}    

\subsection{A posteriori examples}
\textcolor{black}{
In the following examples, we do not impose control invariance on $\mathcal{X}_1$ as required in Assumption \ref{wellpd}. Instead, we enforce the non-control-invariant set $\mathcal{X}$ as the state-constraint set in the numerical MPC problems. Accordingly, the examples should be interpreted as a posteriori trajectory-based evaluations of the computed bounds, rather than as global certificates over the full set $\mathcal{X}$. In the nonlinear example, this is done because constructing a control-invariant set for a general constrained nonlinear system is a complex task, to be studied in future work. In the linear example, we adopt the same setting as in \cite{do2025constraint} to preserve the benchmark used for comparison. For the a posteriori evaluation, the less restrictive set is interpreted as $\mathcal{X}_2=\mathcal{D}$, where $\mathcal{D}$ is selected after the simulations as a compact state domain containing the predicted states generated in the reported trajectories. Since these states already belong to $\mathcal{D}$, imposing the corresponding $\mathcal{X}_2$ constraints would not change the computed optimal trajectories. Thus, the second part of the horizon is effectively unconstrained relative to $\mathcal{D}$.}

\subsubsection{Nonlinear example}
We study the derived bounds on a 6-DoF nonlinear quadrotor model. We adopt the North East Down (NED) orientation, with the inertial frame located on earth's surface, and $O_{xyz}$ fixed at the quadrotor centre of mass.

We model the quadrotor using a simplified nonlinear model valid for small angle deviations \cite{sabatino2015quadrotor}:
\textcolor{black}{
\begingroup
\setlength{\parskip}{0pt}%
\setlength{\abovedisplayskip}{4pt}%
\setlength{\belowdisplayskip}{4pt}%
\setlength{\abovedisplayshortskip}{0pt}%
\setlength{\belowdisplayshortskip}{2pt}%
\setlength{\jot}{1pt}%
\raggedbottom
\begin{equation*}
\small
\begin{aligned}
\dot{\mathbf{x}} &= f(\mathbf{x},\mathbf{u}),\qquad
\mathbf{x} =[x,y,z,\phi,\theta,\psi,u,v,w,p,q,r]^{\top},\\[-0.5mm]
f(\mathbf{x},\mathbf{u}) &= [f_1,\ldots,f_{12}]^{\top},\\[-0.5mm]
f_1 &= w(\phi\psi+\theta)-v(\psi-\phi\theta)+u,\\
f_2 &= v(1+\phi\psi\theta)-w(\phi-\psi\theta)+u\psi,\\
f_3 &= w-u\theta+v\phi,
\qquad f_7 = rv-qw-g\theta,\\
f_4 &= p+r\theta+q\phi\theta,
\qquad f_8 = pw-ru+g\phi,\\
f_5 &= q-r\phi,
\qquad f_9 = qu-pv-\tfrac{\bar{f}_t}{m}, \: \bar{f}_t = f_t - mg\\
f_6 &= r+q\phi,
\qquad f_{10} = \tfrac{I_y-I_z}{I_x}rq+\tfrac{\tau_x}{I_x},\\
f_{11} &= \tfrac{I_z-I_x}{I_y}pr+\tfrac{\tau_y}{I_y},\qquad f_{12} = \tfrac{I_x-I_y}{I_z}pq+\tfrac{\tau_z}{I_z}.
\end{aligned}
\end{equation*}
\endgroup
}

In the above equation $\textbf{x} \in \mathbb{R}^{12}$ is the vector composed of: \textcolor{black}{positions $x,y,z$, angular displacements $\phi,\theta,\psi$, translational velocities $u,v,w$, and angular velocities $p,q,r$ in/around the} directions $o_x,o_y,o_z$, with respect to the inertial frame \cite{sabatino2015quadrotor}. The input vector $\textbf{u}=[\bar{f}_t,\tau_x,\tau_y,\tau_z]^T$, is composed by $\bar{f}_t$, the net thrust force above the UAV's weight and $\tau_x,\tau_y, \tau_z$ are torques generated along the $o_x,o_y,o_z$ directions. Finally, $I_x, I_y, I_z$ are inertia matrix diagonal components (off-diagonal terms neglected).
$\textrm{UAV}_1$ has its initial state $\textbf{x}_{\mathrm{init}} = [1,2,-1,0_{1x9}]^T$, and $\textrm{UAV}_2$ has its state fixed to $\textbf{x}_\mathrm{o} = [0.4,1.5,-0.2,0_{1x9}]^T$. With respect to the NED orientation, in which a negative $Z$ component means displacement above ground. $\textrm{UAV}_1$ starts hovering at $[1,2,-1]^{\top}$ and is to land at the origin while avoiding $\textrm{UAV}_2$, hovering at $[0.4,1.5,-0.2]^{\top}$. We discretize the system with sample time $h=0.4s$ and solve \eqref{eq:VN_Ntilde}. The input and state constraint sets are \textcolor{black}{$\mathcal{U}=\{\textbf{u} \in \mathbb{R}^4| u_{min} \leq \textbf{u}\leq u_{max}\}$} and \textcolor{black}{$\mathcal{X}_2 =\mathcal{D} \subset \mathbb{R}^{12}$}. \textcolor{black}{We impose $\textbf{x}(n+1|\textbf{x}_k)\in\mathcal X$} for all $ n=0,\dots, N-\tilde{N}$ with $\mathcal{X}$ defined in \eqref{eq:state_ctr},
\textcolor{black}{
\begingroup
\begin{equation}
\label{eq:state_ctr}
\small
\begin{aligned}
\mathcal{X}
=
\left\{
\mathbf{x}\in\mathbb{R}^{12}\,\middle|\,
\right.
& \|d_{1,2}\|_2^2 \geq r_{\mathrm{obs}}^2+\epsilon,\quad z\leq 0,\\
& \max\{|\phi|,|\theta|,|\psi|\}\leq \frac{\pi}{9},\\
& \left.
\|v_{1,2}\|_2\leq 2,\quad
\max\{|p|,|q|,|r|\}\leq \frac{\pi}{18}
\right\},
\end{aligned}
\end{equation}
\endgroup
where \(d_{1,2}:=[x-x_o,\ y-y_o,\ z-z_o]^\top\) and
\(v_{1,2}:=[u,\ v,\ w]^\top\). Since $\mathcal{X}$ is unbounded in its position, we use $\mathcal{X}_1 = \mathcal{X} \cap \mathcal{D}$.
}
In \eqref{eq:state_ctr}, we enforce for $\textrm{UAV}_1$ a safety distance around $\textrm{UAV}_2$ of $r_{obs}=0.5$ and set the small tolerance value $\epsilon>0$ to $\epsilon=0.1$. The other constraints are ground avoidance, the small-angle validity region, and bounds on
translational and angular velocities. The stage cost is $l(x(n|x_k),u(n|x_k)) = x(n|x_k)^{\top}Qx(n|x_k) + u(n|x_k)^{\top}Ru(n|x_k)$. 
\textcolor{black}{We test the upper certificate in \eqref{eq:alpha} against the closed-loop performance \(J^{N,\tilde N}_{\infty}(x_k)\), for $N=17$ and $N-\tilde{N}=15$. The parameter \(\alpha\) is obtained explicitly using Theorem \ref{thm1} with $\delta$ being calculated via \eqref{eq:VtildeVdiff1}.}

\textcolor{black}{For the fixed $(N,\tilde{N})$ pair the parameters \(\sigma_1,C_1,\sigma_2,C_2\), and \(\delta\) are computed sequentially, using closed-loop states \(x_j\), \(j=0,\dots,T\), where $T$ is the last sample in the trajectory and \(x_T\) has reached the equilibrium. Then, for each parameter, we choose its most conservative value over the sampled closed-loop trajectory \(x_0,\dots,x_T\) to plug into the $\alpha$ expression on Theorem \ref{thm1}}. \textcolor{black}{The pointwise parameters \(C_1(x_j)\) and \(\sigma_1(x_j)\) associated with Assumption~\ref{ass2} are obtained as
\begin{align}
    \sigma_1(x_j)
    &=
    \min_{n=0,\ldots,N-\tilde N-1}
    \left(
    \frac{\lambda_{N-\tilde N}(x_j)}
    {\lambda_n(x_j)}
    \right)^{\frac{1}{N-\tilde N-n}}, \nonumber \\
    C_1(x_j)
    &=
    \max_{n=0,\ldots,N-\tilde N}
    \frac{\lambda_n(x_j)}
    {\sigma_1(x_j)^n\lambda_0(x_j)} .
\end{align}
}
\begin{align}
    \textcolor{black}{
    \sigma_1
    =
\max_{j=0,\dots,T}\sigma_1(x_j),
    \:
    C_1
    =
    \max_{j=0,\dots,T}C_1(x_j).
    }
\end{align}
\textcolor{black}{
Similarly, for each sampled closed-loop state \(x_j\), we estimate \(\sigma_2(x_j)\) from the auxiliary tail bound \eqref{bunctr} in Assumption~\ref{ass2}. We set \(C_2=1\) and choose \(\sigma_2(x_j)\) as the smallest value in \([0,1)\) satisfying the two instances of \eqref{bunctr} used in the proof of Theorem~\ref{thm1}:
\begin{align*}
V_{\tilde N}(x_s)
&\leq
\lambda_{N-\tilde N}
\sum_{i=1}^{\tilde N}\sigma_2(x_j)^i,
\\
V_1\!\left(x_d^\star(\tilde N-1\mid x_s)\right)
&\leq
\lambda_{N-\tilde N}\sigma_2(x_j)^{\tilde N}.
\end{align*}
At each closed-loop step, we compute the smallest value satisfying the first tail inequality by bisection, while the second inequality gives an explicit lower requirement on \(\sigma_2(x_k)\). We then set \(\sigma_2(x_k)\) as the maximum of the two resulting values, ensuring that both tail inequalities hold at that step. The trajectory-wise parameters used to compute \(\alpha\) in Theorem~\ref{thm1} are then
\[
\sigma_{2}
=
\max_{j=0,\ldots,T}\sigma_2(x_j),
\qquad
C_{2}=1.
\]
}
\textcolor{black}{
To avoid ill-conditioned ratios near the origin, pointwise estimates of $C_1(x_j), \sigma_1(x_j)$ and $\sigma_2(x_j)$ whose denominator costs are below \(10^{-5}\) are discarded. The parameters obtained for the $(N=17, \tilde{N}=2)$ case are listed below 
\begin{align*}
C_1 &= 122.41, & C_2 &= 1.00, & \delta &= 0, \\
\sigma_1 &= 0.43, & \sigma_2 &= 0.85, \\
\alpha &= 0.9997, & V_N^{\tilde N}(x_0) &= 458.3040, \\
J^{N,\tilde N}_{\infty}(x_0) &= 458.3077,
& \frac{V_N^{\tilde N}(x_0)}{\alpha}
&= 458.4339.
\end{align*}
Via \eqref{opt_omega}, the implicit value of \(\omega=0\), indicating that this method does not identify improvement over the naive lower. This is expected given the close agreement between \(V_N^{\widetilde N}(x_0)\) and \(J_\infty^{N,\widetilde N}(x_0)\). Non-trivial lower-bound information is more often extracted for short prediction horizons, where the open-loop cost typically carries insufficient information to reflect the closed-loop cost. In this example, however, the constraints were not chosen to be invariant. Hence, smaller values of \(N\) combined with larger values of \(\tilde N\) often led to impossibility of obtaining any certificate using, for instance, \eqref{eq:VtildeVdiff1} mainly due to infeasibility of \(V^{\tilde N}_{\tilde N}(x_s)\). Given that our existence of $\omega$ is tied to the existence of $\alpha > 0$, this prevented us from using precisely those horizon combinations that could otherwise have yielded a larger value of \(\omega\).
}
\textcolor{black}
{In the same experiment, we have also implemented a traditional MPC, enforcing $\mathcal{X}$ across the whole prediction horizon $N=17$. We then compared its results with those of HC-MPC while keeping $N=17$ and varying $N-\tilde{N} \in \{3,7,11,15\}$. All instances produced simulated trajectories satisfying the safety constraints. For each fixed $N-\tilde{N}$, we recorded  $7$ runs of $100$ steps each. The maximum and average solve time along with the controller's closed loop cost, over all runs for each optimization problem  trajectory is shown on Table \ref{tab:mpc_comparison}. The overall trend, although not monotonic, is a decrease in average solve time as $N-\tilde{N}$ decreases, while the closed-loop cost is non-decreasing.}

\begin{table}[t]
\textcolor{black}{
    \centering
    \caption{HC-MPC vs MPC evaluations.}
    \label{tab:mpc_comparison}
    \resizebox{\columnwidth}{!}{%
        \begin{tabular}{lcccc}
            \toprule
            Controller
            & $N-\widetilde N$
            & Max.\ [ms]
            & Avg.\ [ms]
            & $J_{\mathrm{cl}}$ \\
            \midrule
            HC-MPC & $3$  &  96.314 & 35.298 & 460.501 \\
            HC-MPC & $7$  & 135.491 & 39.717 & 458.308 \\
            HC-MPC & $11$ & 163.233 & 43.003 & 458.308 \\
            HC-MPC & $15$ & 148.353 & 46.129 & 458.308 \\
            MPC    & --   & 195.672 & 47.314 & 458.308 \\
            \bottomrule
        \end{tabular}%
    }
} 
\end{table}

\subsubsection{Linear system comparison}

\begin{figure}[htbp]
\centering
\includegraphics[width=0.43\textwidth]{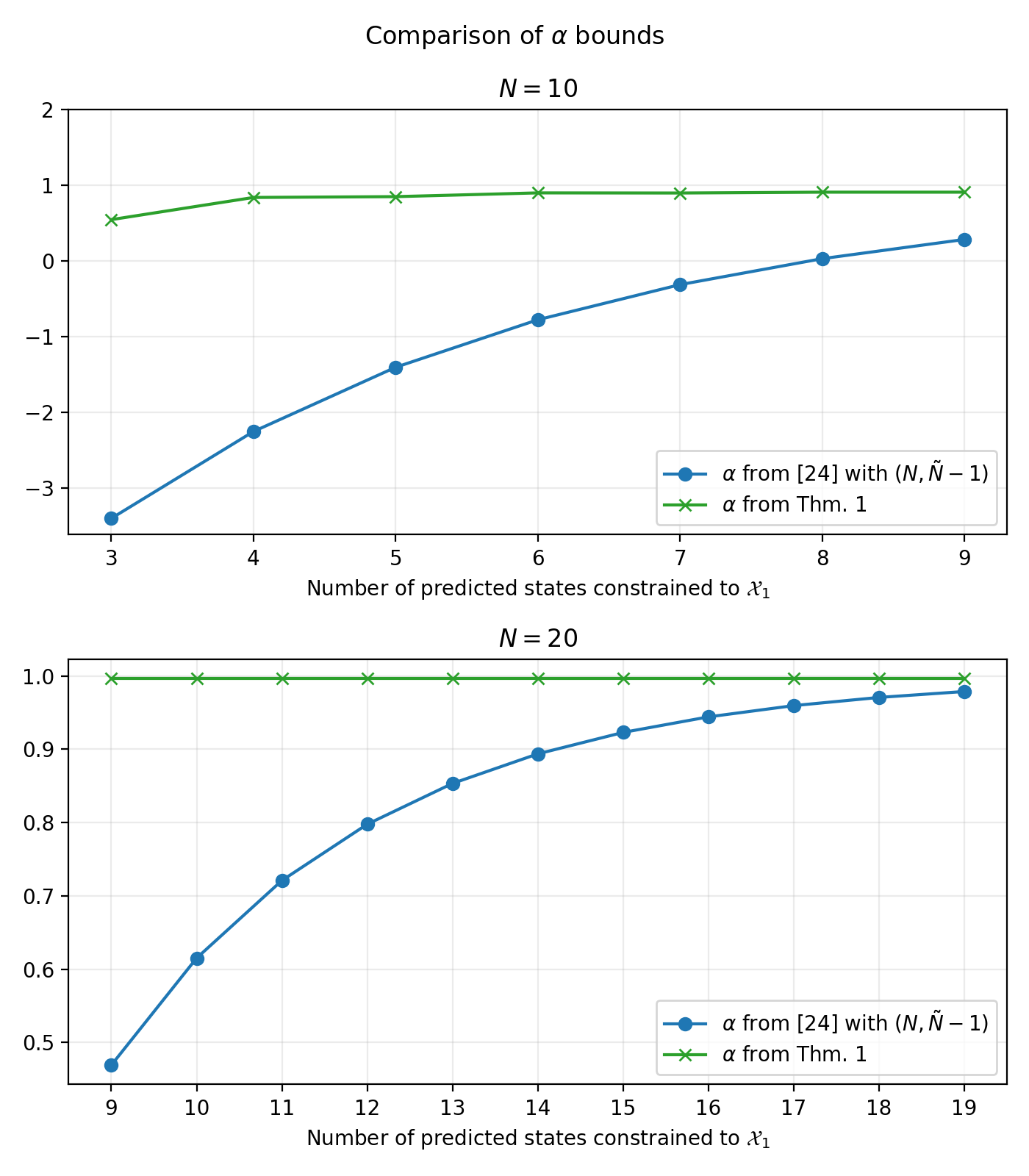}\\
\caption{\label{fig:alp_comp_10_20} 
$\alpha$'s calculation via Theorem 1 and \cite{do2025constraint}
}
\end{figure}

\begin{figure}[htbp]
\centering
\includegraphics[width=0.43\textwidth]{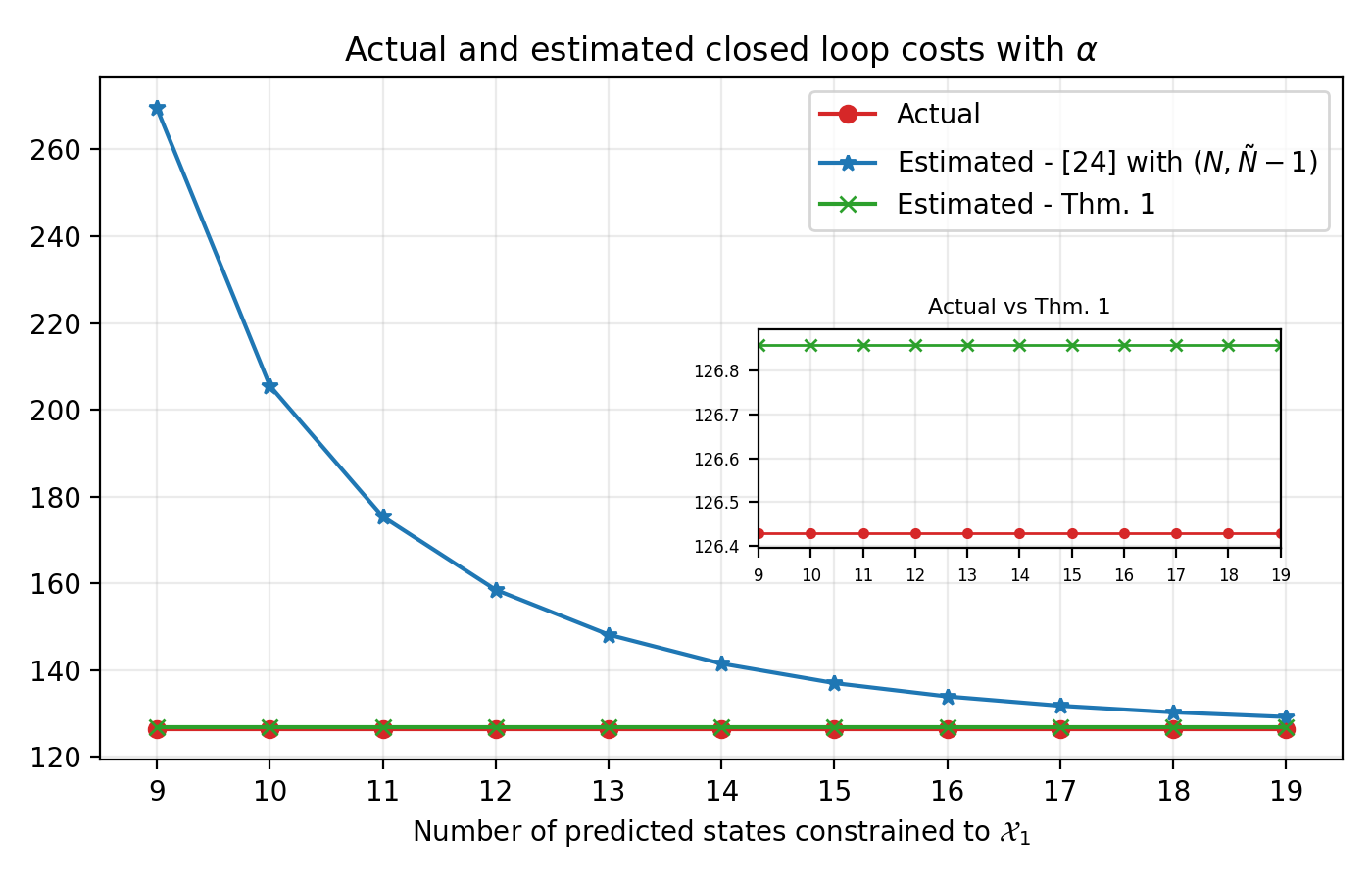}\\
\caption{\label{fig:bd_comp_20} 
\textcolor{black}{Estimated} and closed-loop cost for $N=20$ as a function of \textcolor{black}{the number of predicted states in $\mathcal{X}_1$}.
}
\end{figure}

We compare Theorem \ref{thm1} with \cite{do2025constraint}, which also derives \eqref{eq:alpha} via RDP, but estimates $\alpha$ as:
\begin{equation} \label{eq:alpha_LCSS}
    \alpha = 1-\frac{\beta^{N-\tilde N+1}}{(\beta+1)^{N-\tilde N-1}},
\end{equation}
and $\beta>0$ is obtained for $n\in\{\tilde N+1,\ldots,N\}$ as:
    \begin{subequations}
        \begin{align}
        &V_{\tilde N+1}^{\tilde N}(x_k)\le (\beta+1)V_{\tilde N}^{\tilde N}(x_k), \label{eq:ass1eq1LCSS}\\
            &V_n^{\tilde N}(x_k)\le (\beta+1)l(x_k,u^*_h(N-n|x_k)). \label{eq:ass1eq2LCSS}
        \end{align}
    \end{subequations}
\textcolor{black}{ Following \cite{do2025constraint}, we consider the planar double integrator \(x(i+1|x_k)=Ax(i|x_k)+Bu(i|x_k)\), with
\(x=[p_x,p_y,v_x,v_y]^\top\) collecting positions and velocities, and
\(u=[a_x,a_y]^\top\) collecting accelerations. The stage cost is
\(l(x,u)=x^\top Qx+u^\top Ru\), with input bounds $\mathcal{U}=\{u \in \mathbb{R}^2, s.t. \; -2\leq u_r \leq 2\}$, where $r=\{1,2\}$ are the rows of $u$, and input constraints are enforced for $i=0,\dots, N-1$}. State constraints are: velocity bounds $|v_x(j+1|x_k)|+|v_y(j+1|x_k)| \leq 2$ and for position bounds, we re-utilize the CBF \emph{candidate functions}
\begin{subequations}
    \begin{align}
       & h_1(x(j|x_k)) = \frac{5}{9}p_x(j|x_k)+p_y(j|x_k)+\frac{0.5}{9}, \\
       & h_2(x(j|x_k)) = p_x(j|x_k)-p_y(j|x_k)+1.6,
    \end{align}
\end{subequations}
enforcing constraints via $h_1(x(j+1|x_k))\geq(1-\gamma)h_1(x(j|x_k))$ and $h_2(x(j+1|x_k))\geq(1-\gamma)h_2(x(j|x_k))$, with $\gamma=0.8$ for $j=0,\dots,N-\tilde{N}$. The initial state is set to $x_0=[-0.8,0.6,-0.45,0.65]$, and the target state is the origin. \textcolor{black}{ For comparison, we consider $N=10$ and $N=20$ and repeatedly solve \eqref{eq:VN_Ntilde} while varying $\tilde N$. For each fixed $N$ and each $\tilde N$, following the procedure of the previous example, the parameters required by Theorem \ref{thm1}, using $\delta$ from \eqref{eq:VtildeVdiff1}, are computed over all sampled states $x_k$. As a benchmark, we also compute $\beta$ from the same full trajectory and use it to obtain the corresponding $\alpha$ according to \cite{do2025constraint}. We note that, compared with its counterpart in \cite{do2025constraint}, the formulation in \eqref{eq:VN_Ntilde} enforces one additional predicted state in $\mathcal X_1$. Thus, in set-enforcement terms, a given pair $(N,\tilde N)$ in the present formulation corresponds to $(N,\tilde N-1)$ in \cite{do2025constraint}. Hence, to draw a fair comparison, the results below are presented using the prediction horizon $N$ against the number of predicted states constrained in $\mathcal X_1$ for each method, rather than using our convention so far of $N-\tilde N$. The resulting method comparison for $N=10$ and $N=20$ is shown in Fig. \ref{fig:alp_comp_10_20}.}
The $\alpha$ estimate from Theorem \ref{thm1} via \eqref{eq:VtildeVdiff1} \textcolor{black}{produces higher $\alpha$ estimates and is valid over a larger range of predicted states constrained in $\mathcal{X}_1$ for both $N=10$ and $N=20$, when compared to computations via \eqref{eq:alpha_LCSS}. Up to a certain extent, this is not surprising given that conservativeness enters \eqref{eq:alpha_LCSS}, via the several sufficient conditions used during the induction derivation. For completeness, we also explored Theorem \ref{thm1}'s version using the more conservative holdability-based estimate in \eqref{eq:deltaeq}. This produced the most conservative results among the methods studied, extending $\alpha$'s inapplicability region, and values of $\alpha$ lower than \eqref{eq:alpha_LCSS}. Finally, the closed-loop upper bound and its real value are shown in Fig.~\ref{fig:bd_comp_20}. We use the $\alpha$ values from the bottom graph in Fig.~\ref{fig:alp_comp_10_20}.}     

\section{Conclusion}
\label{sec:conclusion}
\textcolor{black}{
We presented an MPC formulation with two state-constraint sets. The first is control invariant, certifies near-term safety, and ensures recursive feasibility. The second is less restrictive, contains the first, and recovers the partially constrained formulation from \cite{do2025constraint} when \(\mathcal X_2\) is chosen large enough not to restrict the tail predictions. Motivated by receding-horizon safety-critical control, we developed a value-function-difference framework for closed-loop performance analysis. This yields implicit upper- and lower-bound certificates, as
well as horizon-explicit certificates under cost-controllability
assumptions. The lower certificate provides a finite-horizon alternative to the infinite-horizon open-loop value, improving the standard lower bound when specific conditions hold. Linear and nonlinear
examples illustrated the proposed certificates a priori and a posteriori. Future work will address efficient parameter estimation, deepen comparisons across horizon choices, and clarify the relation between the proposed
certificates and existing MPC performance bounds.} 

\appendix
\subsection{Relaxed Dynamic Programming Proofs} \label{rdpproof}
\subsubsection{Proof of Proposition \ref{prop_lb}}

\textcolor{black}{Summing \eqref{eq:dyn_prog_2} from $j=0, \dots, M-1$, and subsequent multiplication by $\alpha$ gives}
\begin{align} \label{eq:sum_bd2}
\alpha \sum^{M-1}_{j=0} l(x_{k+j},\mu^{\tilde{N}}_N(x_{k+j}))
&\coloneq \alpha J^{N,\tilde{N}}_M(x_k) \nonumber \\
&\geq \frac{\alpha}{1-\omega}
\left[ V_N^{\tilde{N}}(x_k)-V_N^{\tilde{N}}(x_{k+M}) \right].
\end{align}
\textcolor{black}{Letting $M\to\infty$ and using Lemma~\ref{lemma_2} yields}
\begin{equation}\label{eq:alpha_lb}
\alpha J_{\infty}^{N,\tilde{N}}(x_k)
\geq
\frac{\alpha}{1-\omega} V_N^{\tilde{N}}(x_k).
\end{equation}
\textcolor{black}{Since \eqref{eq:dyn_prog} holds under the hypotheses of Lemma~\ref{lemma_2}, summing \eqref{eq:dyn_prog} from $j=0, \dots, M-1$ gives}
\begin{align} \label{eq:sum_bd2_upper}
V_N^{\tilde{N}}(x_k)-V_N^{\tilde{N}}(x_{k+M})
\geq
\alpha J^{N,\tilde{N}}_M(x_k).
\end{align}
\textcolor{black}{Again letting $M\to\infty$ and using Lemma~\ref{lemma_2}, we obtain}
\begin{equation}\label{eq:alpha2}
V_N^{\tilde{N}}(x_k)
\geq
\alpha J_{\infty}^{N,\tilde{N}}(x_k)
\geq
\frac{\alpha}{1-\omega} V_N^{\tilde{N}}(x_k),
\end{equation}
for all $x_k\in\mathcal{X}_1$.

\textcolor{black}{Moreover, under the simultaneous validity of \eqref{eq:dyn_prog} and \eqref{eq:dyn_prog_2}, for every $x_k\neq0$ with $\alpha>0$,}
\[
\textcolor{black}{
\alpha
\leq
\frac{
V_N^{\tilde{N}}(x_k)-V_N^{\tilde{N}}(x_{k+1})
}{
l(x_k,\mu_N^{\tilde{N}}(x_k))
}
\leq
1-\omega,}
\]
\textcolor{black}{and hence $\alpha\leq1-\omega$.}

\subsection{Upper bound Proofs} \label{ubproof}

\subsubsection{Proof of Lemma \ref{lem:eqVNtilde}} \textcolor{black}{ The tail of the optimal trajectory of Problem \eqref{eq:VN_Ntilde}, initialized at $x_h^*(N-\tilde N+1|x_k)$, is feasible for the auxiliary problem $V_{\tilde N-1}(x_h^*(N-\tilde N+1|x_k))$. Hence, by optimality, 
\begin{align} 
\sum_{n=N-\tilde N+1}^{N-1} &  l(x_h^*(n|x_k),u_h^*(n|x_k)) \nonumber \\
& \geq V_{\tilde N-1}(x_h^*(N-\tilde N+1|x_k)). 
\end{align}
Conversely, replacing the tail in \eqref{eq:brkdwn} from stage $n=N-\tilde N+1$ onwards, by $V_{\tilde N-1}(x_h^*(N-\tilde N+1|x_k))$ gives a feasible candidate for Problem \eqref{eq:VN_Ntilde}. Since the original trajectory is optimal, the reverse inequality holds. Therefore, 
\begin{align} 
 \sum_{n=N-\tilde N+1}^{N-1} & l(x_h^*(n|x_k),u_h^*(n|x_k)) \nonumber\\
 & = V_{\tilde N-1}(x_h^*(N-\tilde N+1|x_k)). 
\end{align} }

\subsubsection{Proof of Lemma \ref{lem:diff_val_func_ubd}}
\textcolor{black}{Since $x_{k+1}=x_h^*(1|x_k)$, shifting the optimizer of Problem \eqref{eq:VN_Ntilde} at $x_k$ and appending an admissible $\tilde N$-step tail from $x_h^*(N-\tilde N+1|x_k)$ yields a feasible candidate for Problem \eqref{eq:VN_Ntilde} initialized at $x_{k+1}$. Therefore, 
\begin{align} V_N^{\tilde N}(x_{k+1}) & \leq \sum_{n=1}^{N-\tilde N} l(x_h^*(n|x_k),u_h^*(n|x_k)) \nonumber \\
& + V_{\tilde N}^{\tilde N}(x_h^*(N-\tilde N+1|x_k)). 
\end{align} Subtracting this inequality from \eqref{eq:brkdwn}, and replacing the cost accumulated from stage $n=N-\tilde N+1$ onwards by $V_{\tilde N-1}(x_h^*(N-\tilde N+1|x_k))$ according to Lemma~\ref{lem:eqVNtilde}, gives 
\begin{align} 
& V_N^{\tilde N}(x_k)-V_N^{\tilde N}(x_{k+1}) \geq 
\lambda_0 \nonumber \\
& - \Big( V_{\tilde N}^{\tilde N}(x_h^*(N-\tilde N+1|x_k)) - V_{\tilde N-1}(x_h^*(N-\tilde N+1|x_k)) \Big), 
\end{align} which proves the claim. }

\subsubsection{Proof of Lemma \ref{lem:implicit_alpha}}
\textcolor{black}{
By Lemma~\ref{lem:diff_val_func_ubd}, the first inequality in 
\eqref{eq:upbound_dif} holds. For any 
$x_k\in\mathcal X_1\setminus\{0\}$, positive definiteness of 
$l(\cdot,\cdot)$ gives $\lambda_0>0$. Hence, for such $x_k$, the second 
inequality in \eqref{eq:upbound_dif} is equivalent to
\begin{align*}
\alpha
& \leq
1-\frac{1}{\lambda_0}
\left[
V_{\tilde{N}}^{\tilde{N}}(x^*_{h}(N-\tilde{N}+1|x_k)) \right. \\
& \left. -
V_{\tilde{N}-1}(x^*_{h}(N-\tilde{N}+1|x_k))
\right].
\end{align*}
Taking the worst case over $x_k\in\mathcal X_1\setminus\{0\}$ gives 
\eqref{opt_alpha}. The assumed nonnegativity of the intermediate lower bound 
in \eqref{eq:upbound_dif} ensures that the certificate generated by 
\eqref{opt_alpha} is nonnegative. At $x_k=0$, the RDP inequality holds 
trivially. Therefore, if $\alpha\in[0,1]$, then \eqref{eq:upbound_dif} holds 
uniformly over $\mathcal X_1$, which directly implies \eqref{eq:dyn_prog}. 
Lemma~\ref{lem:cl-ub-optimality} then yields the claim.}

\subsubsection{Proof of Lemma \ref{lem:sec_bound}}
\textcolor{black}{
The difference $V_{\tilde N}(x_s)-V_{\tilde N-1}(x_s)$ is bounded by
extending an optimizer of the $(\tilde N-1)$-horizon auxiliary problem by
one feasible step. Specifically, a feasible one-step auxiliary tail from
$x_d^*(\tilde N-1|x_s)$ provides a feasible candidate for
$V_{\tilde N}(x_s)$. Hence, by optimality of $V_{\tilde N}(x_s)$,
\[
V_{\tilde N}(x_s)
\leq
V_{\tilde N-1}(x_s)
+
V_1(x_d^*(\tilde N-1|x_s)).
\]
Therefore, by \eqref{bunctr},
\[
V_{\tilde N}(x_s)-V_{\tilde N-1}(x_s)
\leq
V_1(x_d^*(\tilde N-1|x_s))
\leq
C_2\sigma_2^{\tilde N}\lambda_{N-\tilde N}.
\]
Using \eqref{bctr} to produce 
$\lambda_{N-\tilde{N}} \leq C_1\sigma^{N-\tilde{N}}_1\lambda_0$, proves \eqref{eq:diff_pt2}.
}

\subsubsection{Proof of Theorem\ref{thm1}}
We start by \textcolor{black}{using  \eqref{eq:VtildeVdiff1}}:
  \begin{align*}
    & V_{\tilde{N}}^{\tilde{N}}(x_s) - V_{\tilde{N}}(x_s) \leq \delta \sum^{\tilde{N}-1}_{n=0} l(x^*_{d}(n|x_s),u^*_d(n|x_s))
\end{align*}
Using \eqref{bunctr} on the expression above we obtain:
  \begin{align*}
    & V_{\tilde{N}}^{\tilde{N}}(x_s) - V_{\tilde{N}}(x_s) \leq \textcolor{black}{ C_2 \lambda_{N-\tilde{N}} \left ( \delta \sum^{\tilde{N}}_{n=1} \sigma^n_2 \right)}, \\
     &  V_{\tilde{N}}^{\tilde{N}}(x_s) - V_{\tilde{N}}(x_s) \leq \sigma_2  C_2 \left ( \delta \frac{1-\sigma^{\tilde{N}}_2}{1-\sigma_2} \right) \lambda_{N-\tilde{N}} 
\end{align*}
Using \eqref{bctr}, we write $\lambda_{N-\tilde{N}}$ as a function of $\lambda_0$.
  \begin{align*}
     &  V_{\tilde{N}}^{\tilde{N}}(x_s) - V_{\tilde{N}}(x_s) \leq \sigma_2  C_2  \left ( \delta \frac{1-\sigma^{\tilde{N}}_2}{1-\sigma_2} \right) \lambda_{N-\tilde{N}}  \\
     & \leq C_1\sigma^{N-\tilde{N}}_1 \sigma_2  C_2 \left ( \delta \frac{1-\sigma^{\tilde{N}}_2}{1-\sigma_2} \right) \lambda_0
\end{align*}
Including the second difference in \eqref{eq:diff2}, using \eqref{eq:diff_pt2}, yields:
\begin{align*}
    & V_{\tilde{N}}^{\tilde{N}}(x_s) - V_{\tilde{N}-1}(x_s) \\
    & =  V_{\tilde{N}}^{\tilde{N}}(x_s) - V_{\tilde{N}}(x_s) + V_{\tilde{N}}(x_s) - V_{\tilde{N}-1}(x_s) \\
    & \leq C_1\sigma^{N-\tilde{N}}_1 \sigma_2  C_2 \left ( \delta \frac{1-\sigma^{\tilde{N}}_2}{1-\sigma_2} \right) \lambda_0 +  C_1C_2\sigma^{N-\tilde{N}}_1\sigma^{\tilde{N}}_2 \lambda_0 \\
    & = C_1\sigma^{N-\tilde{N}}_1 \sigma_2  C_2 \left ( \delta \frac{1-\sigma^{\tilde{N}}_2}{1-\sigma_2} + \sigma_2^{\tilde{N}-1} \right) \lambda_0
\end{align*}
Finally, by using \eqref{eq:upbound_dif}, and the inequality above we obtain:
\begin{align} \label{eq:bd_alpha_comp}
& V_N^{\tilde N}(x_k) - V_N^{\tilde N}(f(x_k,\mu^{\tilde{N}}_N(x_k))) \nonumber \\
& \ge \left (1- C_1\sigma^{N-\tilde{N}}_1 \sigma_2  C_2 \left ( \delta \frac{1-\sigma^{\tilde{N}}_2}{1-\sigma_2} + \sigma_2^{\tilde{N}-1} \right) \right) \lambda_0 \nonumber \\
& \geq \alpha \lambda_0
\end{align}
\textcolor{black}{The tightest $\alpha$ based on \eqref{eq:bd_alpha_comp} is:}
\begin{equation}
\alpha = 1- C_1C_2\sigma^{N-\tilde{N}}_1 \sigma_2 \left ( \delta \frac{1-\sigma^{\tilde{N}}_2}{1-\sigma_2} + \sigma_2^{\tilde{N}-1} \right) \nonumber  
\end{equation}

\subsubsection{Proof of Proposition \ref{corol1}}
Analyzing \eqref{eq:alpha} and having an expression for $\alpha$ available, we only need to obtain an upper bound on $ V^{\tilde{N}}_N(x_k)$. A closed expression for it can be obtained using \eqref{eq:brkdwn}. We can then use \eqref{bctr} to bound the first summation in \eqref{eq:brkdwn}. The second summation is equal to $V_{\tilde{N}-1}(x_s)$ via Lemma \ref{lem:eqVNtilde}. Based on this equality we can use \eqref{bunctr} and \eqref{bctr} sequentially in $V_{\tilde{N}-1}(x_s)$ to bound the second summation in \eqref{eq:brkdwn}. This produces
    \begin{align}
       & V^{\tilde{N}}_N(x_k) \leq \sum^{N-\tilde{N}}_{n=0} C_1 \sigma^n_1\lambda_0 + V_{\tilde{N}-1}(x_s) \nonumber\\
       & \leq C_1 \left [ \left(\frac{1-\sigma_1^{N-\tilde{N}+1}}{1-\sigma_1} \right) + \right. \nonumber \nonumber\\ 
   & \left. C_2\sigma_1^{N-\tilde{N}}\sigma_2 \left(\frac{1-\sigma_2^{\tilde{N}-1}}{1-\sigma_2} \right) \right] \lambda_0. 
    \end{align}
Using \eqref{eq:alpha} and $\lambda_0 \leq \max_{u \in \mathcal{U}} l(x_k,u)$ produces \eqref{eq:cl_ub}.

\subsubsection{Proof of Proposition \ref{prop:delta_bd}}
Consider the difference $V_{\tilde{N}}^{\tilde{N}}(x_s) - V_{\tilde{N}}(x_s)$.
\textcolor{black}{If \(x_s=0\), then the zero input yields \(V_{\tilde N}^{\tilde N}(x_s)=V_{\tilde N}(x_s)=0\), so \eqref{eq:VtildeVdiff1} holds trivially and no ratio-based estimate is needed. Hence, we consider $x_s\neq0$. Using the optimal cost of $V_{\tilde N}(x_s)$ and the bridge condition in Proposition \ref{prop:delta_bd}, we obtain}
\begin{align}
    & V^{\tilde{N}}_{\tilde{N}}(x_s) -V_{\tilde{N}}(x_s)
    = V^{\tilde{N}}_{\tilde{N}}(x_s)
    - \sum^{\tilde{N}-1}_{n=0}l(x^*_{d}(n|x_s),u^*_d(n|x_s)) \nonumber\\
    & \leq l(x_s,u_0) + l(z_s,u_1)
    + \sum^{\tilde{N}-1}_{n=2}l(x^*_{d}(n|x_s),u^*_d(n|x_s))  \nonumber\\
    & - \sum^{\tilde{N}-1}_{n=0}l(x^*_{d}(n|x_s),u^*_d(n|x_s)).
\end{align}
\textcolor{black}{The candidate trajectory applies $u_0$ so that $z_s=f(x_s,u_0)\in\mathcal X_1$, then applies $u_1$ so that $f(z_s,u_1)=x_d^*(2|x_s)$, and then follows the tail of the optimizer of $V_{\tilde N}(x_s)$. This trajectory is feasible for $V_{\tilde N}^{\tilde N}(x_s)$ and cancels the tail costs from $n=2$ onward.}
Therefore,
\begin{align}
    & V^{\tilde{N}}_{\tilde{N}}(x_s)-V_{\tilde{N}}(x_s) \nonumber\\
    &\leq
    l(x_s,u_0)-l(x_d^*(0|x_s),u_d^*(0|x_s)) \nonumber\\
    & +l(z_s,u_1)-l(x_d^*(1|x_s),u_d^*(1|x_s)).
\end{align}
\textcolor{black}{For the bridge construction in Proposition \ref{prop:delta_bd}, define the
uniform constant}
\[
\textcolor{black}{
    \delta :=
    \max\left\{
    c-1,\,
    \sup_{\substack{x_k\in\mathcal X_1\setminus\{0\}\\
    \|x_s\|\geq \varepsilon}}
    \left[
    \frac{l(x_s,u_0)+l(z_s,u_1)}
    {\min_{v\in\mathcal U}l(x_d^*(0|x_s),v)}
    -1
    \right]
    \right\}.
}
\]
\textcolor{black}{Since
\(x_d^*(0|x_s)=x_s\neq0\), Assumption~1 gives $\min_{v\in\mathcal U}l(x_d^*(0|x_s),v)>0$. Moreover, \(c\geq1\) implies \(\delta\geq0\). By the local
condition in Proposition \ref{prop:delta_bd} when \(0<\|x_s\|\leq\varepsilon\), and by
the definition of the supremum when \(\|x_s\|\geq\varepsilon\),}
\begin{align}
    & l(x_s,u_0)+l(z_s,u_1)
    -\min_{v\in\mathcal U}l(x_d^*(0|x_s),v) \nonumber \\
    & \leq
    \delta\min_{v\in\mathcal U}l(x_d^*(0|x_s),v).
\end{align}

\textcolor{black}{Therefore, for all
\(x_k\in\mathcal X_1\setminus\{0\}\),}
\begin{align}
    & V^{\tilde{N}}_{\tilde{N}}(x_s)-V_{\tilde{N}}(x_s) \nonumber\\
    &\leq
    l(x_s,u_0)+l(z_s,u_1) \nonumber\\
    &-l(x_d^*(0|x_s),u_d^*(0|x_s))
    -l(x_d^*(1|x_s),u_d^*(1|x_s)) \nonumber\\
    &\leq
    l(x_s,u_0)+l(z_s,u_1)
    -\min_{v\in\mathcal U}l(x_d^*(0|x_s),v) \nonumber\\
    &\leq
    \delta
    \min_{v\in\mathcal U}l(x_d^*(0|x_s),v) \nonumber\\
    &\leq
    \delta
    \sum_{n=0}^{\tilde N-1}
    l(x_d^*(n|x_s),u_d^*(n|x_s)).
\end{align}
\textcolor{black}{Hence, \eqref{eq:VtildeVdiff1} holds uniformly over
\(\mathcal X_1\) with this choice of \(\delta\).}

\textcolor{black}{If the stronger one-step holdability specialization in
Proposition \ref{prop:delta_bd} applies, the bridge is chosen with \(z_s=x_s\). If, in
addition, \(l(x,u)=x^\top Qx+u^\top Ru\), \(Q\succ0\), \(R\succ0\),
and \(0\in\mathcal U\), then}
\[
\textcolor{black}{
    \min_{v\in\mathcal U}l(x_d^*(0|x_s),v)
    =
    x_s^\top Qx_s .
}
\]
\textcolor{black}{Moreover, for \(\|x_s\|\geq\varepsilon\),}
\begin{align}
& \textcolor{black}{
    \frac{l(x_s,u_0)+l(x_s,u_1)}
    {\min_{v\in\mathcal U}l(x_d^*(0|x_s),v)}
    -1
}\nonumber\\
&\textcolor{black}{=
    \frac{2x_s^\top Qx_s+u_0^\top Ru_0+u_1^\top Ru_1}
    {x_s^\top Qx_s}
    -1 \leq
    1+
    \frac{2\max_{u\in\mathcal U}u^\top Ru}
    {\lambda_{\min}(Q)\varepsilon^2}.}
\end{align}
\textcolor{black}{Thus, the same argument gives the uniform choice}
\[
\textcolor{black}{
    \delta
    =
    \max\left\{
    c-1,\,
    1+
    \frac{2\max_{u\in\mathcal U}u^\top Ru}
    {\lambda_{\min}(Q)\varepsilon^2}
    \right\}.
}
\]

\subsubsection{Proof of Proposition \ref{prop_Nstab}}
\textcolor{black}{
Assume that $C_1,C_2>0,\sigma_1,\sigma_2$, and $\delta$ are known and valid
uniformly over all admissible horizon pairs $(N,\tilde N)$ and all
$x_k\in\mathcal X_1$, and let $\sigma=\max\{\sigma_1,\sigma_2\}$. From Theorem \ref{thm1}, $\alpha\geq0 $ is guaranteed if
\begin{align}
    C_1C_2\sigma_1^{N-\tilde N}\sigma_2
    \left(
    \delta\frac{1-\sigma_2^{\tilde N}}{1-\sigma_2}
    +\sigma_2^{\tilde N-1}
    \right)
    \leq 1 .
    \label{eq:stab_ineq_orig}
\end{align}
If $\sigma_1=0$ or $\sigma_2=0$, the left side of
\eqref{eq:stab_ineq_orig} is zero, so $\alpha=1$ and the claim follows
directly. Assume therefore that $\sigma_1,\sigma_2\in(0,1)$, and hence
$\sigma\in(0,1)$. Since $\sigma_1,\sigma_2\leq\sigma$, we have
\begin{align}
    &C_1C_2\sigma_1^{N-\tilde N}\sigma_2
    \left(
    \delta\frac{1-\sigma_2^{\tilde N}}{1-\sigma_2}
    +\sigma_2^{\tilde N-1}
    \right) \nonumber\\
    &\leq C_1C_2\sigma^{N-\tilde N+1}
    \left(
    \delta\frac{1-\sigma_2^{\tilde N}}{1-\sigma_2}
    +\sigma_2^{\tilde N-1}
    \right) \nonumber\\ 
    &<
    C_1C_2\sigma^{N-\tilde N+1}
    \left[
    \frac{\delta}{1-\sigma_2}+1
    \right].
    \label{eq:stab_ineq_1}
\end{align}
In the first inequality we use $\sigma_1,\sigma_2\leq\sigma$, and in the second we use $1-\sigma_2^{\tilde N}<1$ and $\sigma_2^{\tilde N-1}<1$. A sufficient condition for $\alpha\geq0$ is therefore
\begin{align}
    C_1C_2\sigma^{N-\tilde N+1}
    \left[
    \frac{\delta}{1-\sigma_2}+1
    \right]
    \leq 1,
\end{align}
which is guaranteed whenever
\begin{align}
    N-\tilde N+1
    \geq
    \left\lceil
    \frac{
    \log\left(
    C_1C_2\left[\frac{\delta}{1-\sigma_2}+1\right]
    \right)
    }{
    \log\left(\frac{1}{\sigma}\right)
    }
    \right\rceil .
\end{align}
}
\vspace{-0.9cm}
\subsection{Lower-Bound Proofs} \label{lbproof}

\subsubsection{Proof of Lemma \ref{lem:eqVNNtilde}} 

\textcolor{black}{The tail of the optimal trajectory of Problem \eqref{eq:VN_Ntilde}, initialized at $x_h^*(N-\tilde N|x_{k+1})$, is feasible for the auxiliary problem $V_{\tilde N}^{\tilde N}(x_h^*(N-\tilde N|x_{k+1}))$. Hence, by optimality, \begin{align} 
&V_{\tilde N}^{\tilde N}(x_h^*(N-\tilde N|x_{k+1})) \nonumber\\ 
&\leq l(x_h^*(N-\tilde N|x_{k+1}),u_h^*(N-\tilde N|x_{k+1})) \nonumber \\
& + \sum_{n=N-\tilde N+1}^{N-1} l(x_h^*(n|x_{k+1}),u_h^*(n|x_{k+1})). 
\end{align} 
Conversely, replacing the portion of the optimal solution of Problem \eqref{eq:VN_Ntilde} corresponding to stages $n=N-\tilde N,\ldots,N-1$ by $V_{\tilde N}^{\tilde N}(x_h^*(N-\tilde N|x_{k+1}))$ gives a feasible candidate for Problem \eqref{eq:VN_Ntilde}. Since the original trajectory is optimal, the reverse inequality holds. Therefore, 
\begin{align} 
&V_{\tilde N}^{\tilde N}(x_h^*(N-\tilde N|x_{k+1})) \nonumber\\ 
&= l(x_h^*(N-\tilde N|x_{k+1}),u_h^*(N-\tilde N|x_{k+1})) \nonumber \\
& + \sum_{n=N-\tilde N+1}^{N-1} l(x_h^*(n|x_{k+1}),u_h^*(n|x_{k+1})). 
\end{align}}

\subsubsection{Proof of Lemma \ref{lem:diff_val_func_lbd}}

\textcolor{black}{
Since $x_{k+1}=x_h^*(1|x_k)$, applying $u_h^*(0|x_k)$, then following the optimizer of Problem \eqref{eq:VN_Ntilde} initialized at $x_{k+1}$ up to $x_h^*(N-\tilde N|x_{k+1})$, and appending an admissible $(\tilde N-1)$-step tail from $x_h^*(N-\tilde N|x_{k+1})$ yields a feasible candidate for Problem \eqref{eq:VN_Ntilde} initialized at $x_k$. Therefore,
\begin{align}
V_N^{\tilde N}(x_k)
&\leq
\lambda_0
+
\sum_{n=0}^{N-\tilde N-1}
l(x_h^*(n|x_{k+1}),u_h^*(n|x_{k+1})) \nonumber\\
&\quad
+
V_{\tilde N-1}(x_h^*(N-\tilde N|x_{k+1})).
\end{align}
Using Lemma \eqref{lem:eqVNNtilde} to write
\begin{align}
V_N^{\tilde N}(x_{k+1})
&=
\sum_{n=0}^{N-\tilde N-1}
l(x_h^*(n|x_{k+1}),u_h^*(n|x_{k+1})) \nonumber\\
&\quad
+
V_{\tilde N}^{\tilde N}(x_h^*(N-\tilde N|x_{k+1})),
\end{align}
and subtracting gives
\begin{align}
&V_N^{\tilde N}(x_k)-V_N^{\tilde N}(x_{k+1})
\leq
\lambda_0 \nonumber\\
&\quad
-
\Big(
V_{\tilde N}^{\tilde N}(x_h^*(N-\tilde N|x_{k+1}))
-
V_{\tilde N-1}(x_h^*(N-\tilde N|x_{k+1}))
\Big).
\end{align}
}

\subsubsection{Proof of Lemma \ref{lem:implicit_omega}}
\textcolor{black}{
By Lemma~\ref{lem:diff_val_func_lbd}, the first inequality in 
\eqref{eq:upbound_diff_k+1} holds. For any 
$x_k\in\mathcal X_1\setminus\{0\}$, positive definiteness of 
$l(\cdot,\cdot)$ gives $\lambda_0>0$. Hence, for such $x_k$ and 
$\omega<1$, the second inequality in \eqref{eq:upbound_diff_k+1} is 
equivalent to
\begin{align*}
\omega
& \leq
\frac{1}{\lambda_0}
\left[
V_{\tilde{N}}^{\tilde{N}}(x^*_{h}(N-\tilde{N}|x_{k+1})) \right. \\
& \left. -
V_{\tilde{N}-1}(x^*_{h}(N-\tilde{N}|x_{k+1}))
\right].
\end{align*}
Taking the worst case over $x_k\in\mathcal X_1\setminus\{0\}$ gives 
\eqref{opt_omega}. The assumed nonnegativity of the intermediate upper bound 
in \eqref{eq:upbound_diff_k+1}, together with the admissibility condition 
$\omega\in[0,1)$, ensures that the certificate generated by 
\eqref{opt_omega} is meaningful. At $x_k=0$, the lower-RDP inequality holds 
trivially. Therefore, if $\omega\in[0,1)$, then 
\eqref{eq:upbound_diff_k+1} holds uniformly over $\mathcal X_1$, which 
directly implies \eqref{eq:dyn_prog_2}.}

\subsubsection{Proof of Theorem \ref{thm2}}
\textcolor{black}{We first lower bound the difference to the right of the inequality in \eqref{eq:diff3}. Let $(x_d^*(\cdot|x_p),u_d^*(\cdot|x_p))$ be a selected state-input optimizer pair of the $\tilde N$-horizon auxiliary problem $V_{\tilde N}(x_p)$. The truncated trajectory obtained by retaining the first $\tilde N-1$ stages of this optimizer is feasible for the $(\tilde N-1)$-horizon auxiliary problem. Therefore, by optimality of $V_{\tilde N-1}(x_p)$,}
\begin{align*}
V_{\tilde N-1}(x_p)
&\leq
\sum_{n=0}^{\tilde N-2}
l(x^*_{d}(n|x_p),u^*d(n|x_p)).
\end{align*}
\textcolor{black}{On the other hand, by definition of the same $\tilde N$-horizon optimizer,}
\begin{align*}
V_{\tilde N}(x_p)
&=
\sum_{n=0}^{\tilde N-1}
l(x^*_{d}(n|x_p),u^*_d(n|x_p)).
\end{align*}
\textcolor{black}{Subtracting the previous two relations yields}
\begin{align*}
V_{\tilde N}(x_p)-V_{\tilde N-1}(x_p)
&\geq
l(x^*_{d}(\tilde N-1|x_p),
u^*_d(\tilde N-1|x_p)).
\end{align*}
\textcolor{black}{Using \eqref{blctr} with $n=\tilde N-1$ gives}
\begin{align*}
V_{\tilde N}(x_p)-V_{\tilde N-1}(x_p)
&\geq
C_4\sigma_4^{\tilde N}\lambda^+_{N-\tilde N}.
\end{align*}
\textcolor{black}{Applying \eqref{lctr} with $n=N-\tilde N$, and then using $\lambda_0^+\geq \kappa\lambda_0$, yields}
\begin{align} \label{eq:diff_pt4}
V_{\tilde{N}}(x_p) - V_{\tilde{N}-1}(x_p)
&\geq
C_3C_4\sigma^{N-\tilde{N}}_3
\sigma^{\tilde{N}}_4 \lambda^+_0 \nonumber \\
&\geq
C_3C_4\kappa\sigma^{N-\tilde{N}}_3
\sigma^{\tilde{N}}_4 \lambda_0 .
\end{align}
\textcolor{black}{Moreover, since $V_{\tilde N}^{\tilde N}(x_p)\geq V_{\tilde N}(x_p)$, we obtain}
\begin{align*}
V_{\tilde N}^{\tilde N}(x_p)-V_{\tilde N-1}(x_p)
&\geq
C_3C_4\kappa\sigma^{N-\tilde{N}}_3
\sigma^{\tilde{N}}_4 \lambda_0 .
\end{align*}
\textcolor{black}{Using this estimate in the decrement bound of Lemma \ref{lem:diff_val_func_lbd} yields}
\begin{align}\label{eq:omega_bound_comp}
&V_N^{\tilde N}(x_k) - V_N^{\tilde N}(f(x_k,\mu_N^{\tilde N}(x_k))) \nonumber \\
&\leq \lambda_0 -
\left(
V_{\tilde N}^{\tilde N}(x_p)-V_{\tilde N-1}(x_p)
\right) \nonumber \\
&\leq \left[
1-
C_3C_4\kappa\sigma_3^{N-\tilde N}\sigma_4^{\tilde N}
\right]\lambda_0 \leq (1-\omega)\lambda_0.
\end{align}
\textcolor{black}{Therefore, since $\lambda_0=l(x_k,\mu_N^{\tilde N}(x_k))$, \eqref{eq:omega_bound_comp} establishes the lower-RDP condition \eqref{eq:dyn_prog_2}, and Proposition~\ref{prop_lb} yields the lower bound \eqref{eq:partial_lb}. Thus, the tightest $\omega$ based on \eqref{eq:omega_bound_comp} is}
\begin{equation*}
\omega =
C_3 C_4 \kappa \sigma^{N-\tilde{N}}_3 \sigma_4^{\tilde{N}}.
\end{equation*}
\textcolor{black}{provided that this value is admissible, namely $\omega\in[0,1)$.}

\subsubsection{Proof of Proposition \ref{corol2}}

\textcolor{black}{By definition,
\begin{equation}
J_{\infty}^{N,\tilde N}(x_k)
=
\lambda_0+J_{\infty}^{N,\tilde N}(x_{k+1})
\geq J_{\infty}^{N,\tilde N}(x_{k+1}).
\end{equation}
}
\textcolor{black}{Using \eqref{eq:partial_lb} at \(x_{k+1}\), it remains to lower bound
\(V_N^{\tilde N}(x_{k+1})\). Let \(x_p=x_h^*(N-\tilde N|x_{k+1})\). By
Lemma~\ref{lem:eqVNNtilde},}
\begin{equation}
V_N^{\tilde N}(x_{k+1})
=
\sum_{n=0}^{N-\tilde N-1}
l(x_h^*(n|x_{k+1}),u_h^*(n|x_{k+1}))
+
V_{\tilde N}^{\tilde N}(x_p).
\end{equation}
\textcolor{black}{Since \(V_{\tilde N}^{\tilde N}(x_p)\geq V_{\tilde N}(x_p)\), using the lower-envelope bounds in \eqref{lctr} and \eqref{blctr} gives}
\begin{align}
& V_N^{\tilde N}(x_{k+1})
\geq
C_3\left(\frac{1-\sigma_3^{N-\tilde N}}{1-\sigma_3}\right)\lambda_0^+
+
V_{\tilde N}(x_p) \nonumber\\
&\geq
C_3\left[
\left(\frac{1-\sigma_3^{N-\tilde N}}{1-\sigma_3}\right)
+
C_4\sigma_3^{N-\tilde N}\sigma_4
\left(\frac{1-\sigma_4^{\tilde N}}{1-\sigma_4}\right)
\right]\lambda_0^+ .
\end{align}
\textcolor{black}{Finally, using \eqref{eq:partial_lb}, 
\(\lambda_0^+\geq\kappa\lambda_0\), and
\(\lambda_0\geq\min_{u\in\mathcal U}l(x_k,u)\), results in
\eqref{eq:cl_lb}.}
\vspace{-0.4cm}
\section*{References}
\vspace{-0.7cm}
\bibliographystyle{IEEEtran}
\bibliography{refs}

@book{grune2017nonlinear,
  title={Nonlinear model predictive control},
  author={Gr{\"u}ne, Lars and Pannek, J{\"u}rgen},
  year={2017},
  publisher={Springer}
}

@article{grune2008infinite,
  title={On the infinite horizon performance of receding horizon controllers},
  author={Gr{\"u}ne, Lars and Rantzer, Anders},
  journal={IEEE Transactions on Automatic Control},
  volume={53},
  number={9},
  pages={2100--2111},
  year={2008},
  publisher={IEEE}
}

@inproceedings{bemporad2000explicit,
  title={The explicit solution of model predictive control via multiparametric quadratic programming},
  author={Bemporad, Alberto and Morari, Manfred and Dua, Vivek and Pistikopoulos, Efstratios N},
  booktitle={Proceedings of the 2000 American control conference. ACC (IEEE Cat. No. 00CH36334)},
  volume={2},
  pages={872--876},
  year={2000},
  organization={IEEE}
}

@article{grune2010analysis,
  author  = {Gr{\"u}ne, Lars and Pannek, J{\"u}rgen and Seehafer, Martin and Worthmann, Karl},
  title   = {Analysis of unconstrained nonlinear {MPC} schemes with time varying control horizon},
  journal = {SIAM Journal on Control and Optimization},
  volume  = {48},
  number  = {8},
  pages   = {4938--4962},
  year    = {2010},
  doi     = {10.1137/090758696}
}

@ARTICLE{do2025constraint,
  author={Andre Do Nascimento, Allan and Wang, Han and Papachristodoulou, Antonis and Margellos, Kostas},
  journal={IEEE Control Systems Letters}, 
  title={Constraint Horizon in Model Predictive Control}, 
  year={2025},
  volume={9},
  number={},
  pages={1676-1681},
  doi={10.1109/LCSYS.2025.3580355}}

@misc{sabatino2015quadrotor,
  title={Quadrotor control: modeling, nonlinearcontrol design, and simulation},
  author={Sabatino, Francesco},
  year={2015}
}

@article{shamma1997linear,
  title={Linear nonquadratic optimal control},
  author={Shamma, Jeff S and Xiong, Dapeng},
  journal={IEEE Transactions on Automatic Control},
  volume={42},
  number={6},
  pages={875--879},
  year={1997},
  publisher={IEEE}
}

@article{grune2009practical,
  title={Practical {NMPC} suboptimality estimates along trajectories},
  author={Gr{\"u}ne, Lars and Pannek, J{\"u}rgen},
  journal={Systems \& Control Letters},
  volume={58},
  number={3},
  pages={161--168},
  year={2009},
  publisher={Elsevier}
}

@phdthesis{worthmann2011stability,
  title={Stability analysis of unconstrained receding horizon control schemes},
  author={Worthmann, Karl},
  year={2011}
}

@article{grune2013economic,
  title={Economic receding horizon control without terminal constraints},
  author={Gr{\"u}ne, Lars},
  journal={Automatica},
  volume={49},
  number={3},
  pages={725--734},
  year={2013},
  publisher={Elsevier}
}

@article{faulwasser2018economic,
  title={Economic nonlinear model predictive control},
  author={Faulwasser, Timm and Gr{\"u}ne, Lars and M{\"u}ller, Matthias A and others},
  journal={Foundations and Trends{\textregistered} in Systems and Control},
  volume={5},
  number={1},
  pages={1--98},
  year={2018},
  publisher={Now Publishers, Inc.}
}

@inproceedings{grune2007computing,
  title={Computing stability and performance bounds for unconstrained NMPC schemes},
  author={Gr{\"u}ne, Lars},
  booktitle={2007 46th IEEE Conference on Decision and Control},
  pages={1263--1268},
  year={2007},
  organization={IEEE}
}

@book{grne2013nonlinear,
  title={Nonlinear model predictive control: theory and algorithms},
  author={Gr{\"u}ne, Lars and Pannek, J{\"u}rgen},
  year={2013},
  publisher={Springer Publishing Company, Incorporated}
}

@article{findeisen2011robustness,
  title={Robustness of prediction based delay compensation for nonlinear systems},
  author={Findeisen, Rolf and Gr{\"u}ne, Lars and Pannek, J{\"u}rgen and Varutti, Paolo},
  journal={IFAC Proceedings Volumes},
  volume={44},
  number={1},
  pages={203--208},
  year={2011},
  publisher={Elsevier}
}

@article{wabersich2021predictive,
  title={A predictive safety filter for learning-based control of constrained nonlinear dynamical systems},
  author={Wabersich, Kim Peter and Zeilinger, Melanie N},
  journal={Automatica},
  volume={129},
  pages={109597},
  year={2021},
  publisher={Elsevier}
}

@article{wieland2007constructive,
  title={Constructive safety using control barrier functions},
  author={Wieland, Peter and Allg{\"o}wer, Frank},
  journal={IFAC Proceedings Volumes},
  volume={40},
  number={12},
  pages={462--467},
  year={2007},
  publisher={Elsevier}
}

@inproceedings{ames2019control,
  title={Control barrier functions: Theory and applications},
  author={Ames, Aaron D and Coogan, Samuel and Egerstedt, Magnus and Notomista, Gennaro and Sreenath, Koushil and Tabuada, Paulo},
  booktitle={2019 18th European control conference (ECC)},
  pages={3420--3431},
  year={2019},
  organization={Ieee}
}

@inproceedings{zeng2021safety,
  title={Safety-critical model predictive control with discrete-time control barrier function},
  author={Zeng, Jun and Zhang, Bike and Sreenath, Koushil},
  booktitle={2021 American Control Conference (ACC)},
  pages={3882--3889},
  year={2021},
  organization={IEEE}
}

@inproceedings{do2023game,
  title={A game theoretic approach for safe and distributed control of unmanned aerial vehicles},
  author={do Nascimento, Allan Andre and Papachristodoulou, Antonis and Margellos, Kostas},
  booktitle={2023 62nd IEEE Conference on Decision and Control (CDC)},
  pages={1070--1075},
  year={2023},
  organization={IEEE}
}

@inproceedings{do2024probabilistically,
  title={Probabilistically safe controllers based on control barrier functions and scenario model predictive control},
  author={Do Nascimento, Allan Andre and Papachristodoulou, Antonis and Margellos, Kostas},
  booktitle={2024 IEEE 63rd Conference on Decision and Control (CDC)},
  pages={1814--1819},
  year={2024},
  organization={IEEE}
}

@inproceedings{sabouni2024reinforcement,
  title={Reinforcement learning-based receding horizon control using adaptive control barrier functions for safety-critical systems},
  author={Sabouni, Ehsan and Ahmad, HM Sabbir and Giammarino, Vittorio and Cassandras, Christos G and Paschalidis, Ioannis Ch and Li, Wenchao},
  booktitle={2024 IEEE 63rd Conference on Decision and Control (CDC)},
  pages={401--406},
  year={2024},
  organization={IEEE}
}

@article{blanchini1999set,
  title={Set invariance in control},
  author={Blanchini, Franco},
  journal={Automatica},
  volume={35},
  number={11},
  pages={1747--1767},
  year={1999},
  publisher={Elsevier}
}

@inproceedings{grandia2021multi,
  title={Multi-layered safety for legged robots via control barrier functions and model predictive control},
  author={Grandia, Ruben and Taylor, Andrew J and Ames, Aaron D and Hutter, Marco},
  booktitle={2021 IEEE International Conference on Robotics and Automation (ICRA)},
  pages={8352--8358},
  year={2021},
  organization={IEEE}
}

@article{lorenzen2019stochastic,
  title={Stochastic model predictive control without terminal constraints},
  author={Lorenzen, Matthias and M{\"u}ller, Matthias A and Allg{\"o}wer, Frank},
  journal={International Journal of Robust and Nonlinear Control},
  volume={29},
  number={15},
  pages={4987--5001},
  year={2019},
  publisher={Wiley Online Library}
}

@incollection{grune2018dynamic,
  title={Dynamic programming, optimal control and model predictive control},
  author={Gr{\"u}ne, Lars},
  booktitle={Handbook of model predictive control},
  pages={29--52},
  year={2018},
  publisher={Springer}
}

@article{worthmann2014role,
  title={The role of sampling for stability and performance in unconstrained nonlinear model predictive control},
  author={Worthmann, Karl and Reble, Marcus and Grune, Lars and Allgower, Frank},
  journal={SIAM Journal on Control and Optimization},
  volume={52},
  number={1},
  pages={581--605},
  year={2014},
  publisher={SIAM}
}

@article{limon2006stability,
  title={On the stability of constrained MPC without terminal constraint},
  author={Lim{\'o}n, Daniel and Alamo, Teodoro and Salas, Francisco and Camacho, Eduardo F},
  journal={IEEE transactions on automatic control},
  volume={51},
  number={5},
  pages={832--836},
  year={2006},
  publisher={IEEE}
}

@article{boccia2014stability,
  title={Stability and feasibility of state constrained MPC without stabilizing terminal constraints},
  author={Boccia, Andrea and Gr{\"u}ne, Lars and Worthmann, Karl},
  journal={Systems \& control letters},
  volume={72},
  pages={14--21},
  year={2014},
  publisher={Elsevier}
}

@article{grimm2005model,
  title={Model predictive control: For want of a local control Lyapunov function, all is not lost},
  author={Grimm, Gene and Messina, Michael J. and Tuna, Sezai E. and Teel, Andrew R.},
  journal={IEEE Transactions on Automatic Control},
  volume={50},
  number={5},
  pages={546--558},
  year={2005},
  publisher={IEEE},
  doi={10.1109/TAC.2005.847055}
}

@inproceedings{tuna2006shorter,
  title={Shorter horizons for model predictive control},
  author={Tuna, Sezai E. and Messina, Michael J. and Teel, Andrew R.},
  booktitle={Proceedings of the American Control Conference},
  pages={863--868},
  year={2006},
  organization={IEEE}
}

@article{grimm2007nominally,
  title={Nominally robust model predictive control with state constraints},
  author={Grimm, Gene and Messina, Michael J. and Tuna, Sezai E. and Teel, Andrew R.},
  journal={IEEE Transactions on Automatic Control},
  volume={52},
  number={10},
  pages={1856--1870},
  year={2007},
  publisher={IEEE},
  doi={10.1109/TAC.2007.906187}
}

@article{grune2009analysis,
  title={Analysis and design of unconstrained nonlinear MPC schemes for finite and infinite dimensional systems},
  author={Gr{\"u}ne, Lars},
  journal={SIAM Journal on Control and Optimization},
  volume={48},
  number={2},
  pages={1206--1228},
  year={2009},
  publisher={SIAM}
}
\vspace{-1.5cm}

\begin{IEEEbiographynophoto}
{Allan Andre do Nascimento} received the B.Sc. degree (\emph{cum laude}) in Mechanical Engineering from the Federal University of Rio de Janeiro, Brazil, in 2013, the M.Sc. degree (\emph{distinction}) in Robotics from King's College London, United Kingdom, in 2016, and the M.Sc. degree in Electrical Engineering from the Royal Institute of Technology (KTH), Sweden, in 2019. He is currently a D.Phil. candidate at the University of Oxford. His research interests 
include nonlinear and optimal control, safe control and optimization with applications
to robotic systems.
\end{IEEEbiographynophoto}
\vspace{-1.0cm}
\begin{IEEEbiographynophoto}
{Han Wang} received the B.S. degree in Information Security from Shanghai Jiao Tong University, China, in 2020, and the Ph.D. degree in control from the University of Oxford, UK, in 2024. He is currently a Postdoctoral Researcher at ETH Zürich. His research interests include safe and stable control design, learning to control and convex optimization.
\end{IEEEbiographynophoto}
\vspace{-1.0cm}
\begin{IEEEbiographynophoto}
{Antonis Papachristodoulou}(Fellow, IEEE) received the M.A./M.Eng. degree in Electrical and Information Sciences from the University of Cambridge, United Kingdom in 2000, and the Ph.D. degree in Control and Dynamical Systems (with a minor in Aeronautics) from
the California Institute of Technology, Pasadena, CA,
USA in 2004. He joined the University of Oxford, United Kingdom as faculty in 2005, where he is currently the Professor of Control Engineering. His research interests include large-scale nonlinear systems analysis, sum-of-squares programming, synthetic and systems biology, networked systems, and flow control.
\end{IEEEbiographynophoto}
\vspace{-1cm}
\begin{IEEEbiographynophoto}
{Kostas Margellos}(Senior member, IEEE) received the Diploma degree in Electrical Engineering from the University of Patras, Greece, in 2008, and the Ph.D. degree in Control Engineering from ETH Zürich, Switzerland, in 2012. From 2013 to 2015, he was a Postdoctoral Researcher with ETH Zürich, UC Berkeley, USA, and Politecnico di Milano, Italy. In 2016, he joined the Control Group, Department of Engineering Science, University of Oxford, United
Kingdom, where he is currently an Associate Professor. His research interests include optimization and control of complex uncertain systems, with applications to energy and transportation networks.
\end{IEEEbiographynophoto}
\end{document}